\documentclass[aps, pre, amsmath, amssymb, reprint]{revtex4-2}
\usepackage{commath}
\usepackage{graphicx}
\usepackage{dcolumn}
\usepackage{bbm}
\usepackage[utf8]{inputenc}
\usepackage[T1]{fontenc}
\usepackage{mathptmx}
\usepackage{etoolbox}
\usepackage{float}
\usepackage[linesnumbered,ruled]{algorithm2e}
\usepackage{stmaryrd}
\usepackage{mathptmx}
\usepackage{lipsum}

\DeclareMathOperator*{\bigtimes}{\vartimes}
\DeclareMathOperator*{\argmin}{arg\,min}

\DeclareRobustCommand{\rchi}{{\mathpalette\irchi\relax}}
\newcommand{\irchi}[2]{\raisebox{\depth}{$#1\chi$}} 
\newcommand{\olsi}[1]{\,\overline{\!{#1}}} 

\def\doubleunderline#1{\underline{\underline{#1}}}

\makeatletter
\def\@email#1#2{%
 \endgroup
 \patchcmd{\titleblock@produce}
  {\frontmatter@RRAPformat}
  {\frontmatter@RRAPformat{\produce@RRAP{*#1\href{mailto:#2}{#2}}}\frontmatter@RRAPformat}
  {}{}
}%
\makeatother
\begin{document}

\title{A directed walk in probability space that locates mean field solutions to spin models}
\author{Yizhi Shen}
\author{Adam P. Willard}
\email{awillard@mit.edu}
\affiliation{Department of Chemistry, Massachusetts Institute of Technology, Cambridge, Massachusetts 02139, USA}
\date{\today}

\begin{abstract}
Despite their formal simplicity, most lattice spin models cannot be easily solved, even under the simplifying assumptions of mean field theory. 
In this manuscript, we present a method for generating mean field solutions to classical continuous spins. 
We focus our attention on systems with non-local interactions and non-periodic boundaries, which require careful handling with existing approaches, such as Monte Carlo sampling. Our approach utilizes functional optimization to derive a closed-form optimality condition and arrive at self-consistent mean field equations. We show that this approach significantly outperforms conventional Monte Carlo sampling in convergence speed and accuracy.
To convey the general concept behind the approach, we first demonstrate its application to a simple system - a finite one-dimensional dipolar chain in an external electric field. 
We then describe how the approach naturally extends to more complicated spin systems and to continuum field theories. Furthermore, we numerically illustrate the efficacy of our approach by highlighting its utility on nonperiodic spin models of various dimensionality.
\end{abstract}

\maketitle

\section{\label{sec:Intro} Introduction}
Classical spin models constitute a ubiquitous tool in physical sciences due to their abilities to describe the features and phase behaviors of a wide variety of different physical systems. 
Many properties of these systems can be efficiently understood by applying the mean field (MF) approximation, where each spin is assumed to interact with a static environment determined self-consistently to represent the average. 
In isotropic systems, such as the periodically replicated Ising model, all spins are statistically identical. 
The mean field solution can therefore be captured within a single self-consistent equation. 
By contrast, anisotropic systems, such as those containing an interface, give rise to a more complicated multidimensional array of equations. 
When formal solutions to these equations are inaccessible, for example by their transcendental nature, they can be evaluated algorithmically with adaptive sampling schemes. 
Even in the mean field case, however, typical sampling schemes involve a stochastic walk of some sort in the system configuration space, which becomes prone to frustration as the system size and heterogeneity increases.

In this manuscript, we eliminate such frustration by establishing, instead, a deterministic walk in the space of configurational probabilities.
As we demonstrate, this approach converges rapidly and can be designed to target the correct mean field solution.
Our approach utilizes the joint framework of directional statistics and variational optimization.
The underlying information-theoretic perspective paves a contextualized and accelerated way to study the properties of a wider class of spin models encountered in statistical mechanics, theoretical neurobiology, and artificial intelligence.

Spin systems are capable of modelling specific functionalities and revealing fundamental insights for diverse phenomena that occur in condensed phase systems.~\cite{AnisotropicSpin1,AnisotropicSpin2,AnisotropicSpin3,AnisotropicSpin4} For instance, a monolayer rotor model provides a mechanistic explanation of magnetization reversal observed in magnetic materials~\cite{MagnetReversal1,MagnetReversal2}, while water diffusion and proton conduction in nanopores can be described by a discrete charge model on a segmented lattice~\cite{water1,water2,water3}. In practice, it is nearly impossible to solve a designated spin model exactly, except in a handful of pedagogical cases~\cite{chandler1987introduction,Baxter:1982zz} where the system equilibrium weight can be decomposed into computable subsystem weights according to a simple spin connection topology. When such a decomposition is unavailable, an approximate decomposition can be accomplished via mean field theory, which serves as a zeroth order approximation to the exact theory. 

\begin{figure}[H]
\centering
\includegraphics[width=.48\textwidth]{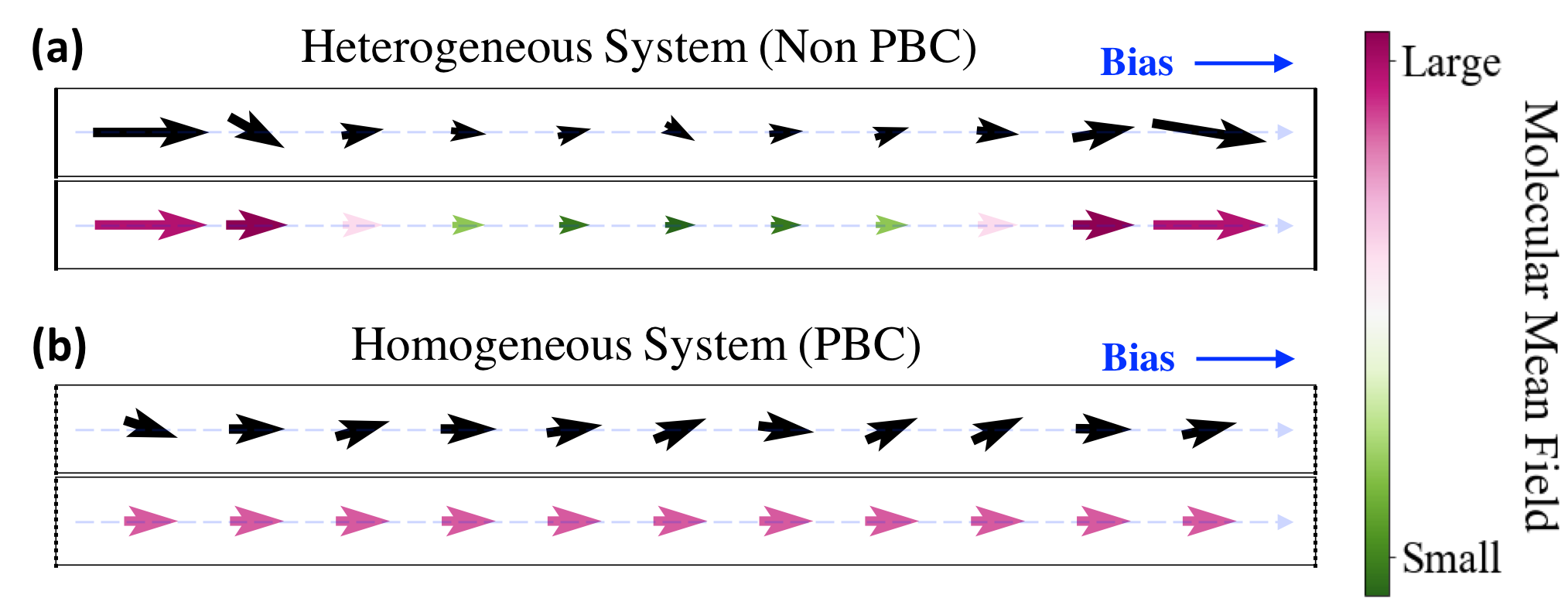}
\caption{
A schematic contrast of mean field theory applied to homogeneous and heterogeneous systems.
Each panel contains a one-dimensional array of dipoles subject to a uniform field biasing the spins to the right, where the top section (black arrows) and bottom section (colored arrows) depict a typical equilibrium configuration and the mean field solution, respectively. 
Panel $(a)$ shows a spatially heterogeneous system with open boundary conditions and panel $(b)$ shows a spatially homogeneous system with periodic boundary conditions.
Spatial heterogeneity leads to larger dipolar response at the system boundaries, an illustration of dipolar screening.
} 
\label{fig:intro}
\end{figure}

Notably, the complexity of a given spin model is set by the length scale of spin-spin interactions. 
When the interactions remain highly local, as in the standard nearest-neighbor Ising model, the majority of spins in the system can be treated as statistically independent. 
The resulting anisotropic mean field equations are hence weakly coupled. For long-range interactions, such as those occurring between the charged species, the growing cross-dependence of these coupled equations poses a challenge for obtaining exact solutions, even with statistical correlations removed by the mean field assumptions. Figure~\ref{fig:intro} highlights this additional complexity from anisotropy and
long-range interactions: the former gives rise to a unique mean field equation for each lattice site and the latter makes the spin topology computationally irreducible.
In principle, one may overcome the associated challenge by employing stochastic sampling methods that bias towards the thermodynamically relevant system configurations. Unfortunately, such methods can fail to converge, especially in cases where the number of metastable basins in the energy landscape proliferates.~\cite{MCtrap,MCtrap2}

In this manuscript we focus our attention on the mean field analysis of finite heterogeneous spin systems. 
In particular, we utilize the well-known Gibbs-Bogoliubov-Feynman variational inequality~\cite{chandler1987introduction,GBF,GBF2} and reframe the mean field approximation as a constrained optimization on the free energy functional of configurational probabilities. We reexpress the infinite-dimensional optimization as an equivalent finite-dimensional optimization over a compact set in the Euclidean space that contains the mean spin characters. The latter problem can be easily solved recursively. We show that this procedure leads to a class of self-consistent mean field equations that are otherwise difficult to derive analytically, and we prove that the iterative procedure is guaranteed to converge. 

Our approach invokes the concept of Markov random fields, also known as factor graphs.~\cite{MarkovRandomField}
This generalized variant of lattice model prescribes adjacency relations between a cloud of random variables, as edges between nodes of a graph, based on their conditional dependence.
Modeling physical spin systems using graphical networks is practically advantageous due to the factorizability native to the accessible system degrees of freedom. That is, the joint distribution of a cluster of variables (\textit{e.g.}, spin configurations) can be factorized as a product over elemental weighting functions (\textit{e.g.}, potential energy surface), each of which involves relatively few variables. Recent progress has been made on the theoretical and practical aspects of Ising model as factor graph through the variational perspectives.~\cite{DemboMontanari,Koehler2018,Koehler2019} Here we demonstrate the ubiquity of fast convergence associated with the mean field iterative approach.  
For illustrative purposes, we examine a trivial representative model composed of a finite chain of freely rotating dipoles.
Even this simple model yields a multimodal energy landscape that can frustrate standard methodologies.
Despite the simplicity of this illustrative model, we note that the same analysis can be applied to a wider range of more complex spin models, as we discuss later in our manuscript.

The manuscript is organized as follows. In Section \ref{sec:model}, we introduce the one-dimensional dipolar lattice as a minimal model with a smooth and continuous multimodal energy landscape. Section \ref{sec:MFT} briefly reviews the mean field formalism in the information-theoretic context. In Section \ref{sec:main result}, we simply state, without proofs, the form and applicability of our mean field iterator. Section \ref{sec:iterator} entails a derivation of the iterator for solving the dipolar model and presents approximation theorems for which relevant model parameters control the iteration convergence. We then address spin models with more general state spaces and interactions in Section \ref{sec:genearlize}, and expand the analogies to include continuum field theories. In Section \ref{sec:cost}, we numerically justify the efficiency and stability of the method outlined in Sections \ref{sec:iterator} and \ref{sec:genearlize} by examining the mean field convergence for various systems, ranging from the one-dimensional dipolar chain to the three-dimensional Heisenberg slab.

\section{\label{sec:model} Model Description}

\subsection{\label{sec:model restriction} Restriction on model space}
We first point out that the concept presented within the work is not model specific, \textit{i.e.}, we will be able to systematically generate directed walks in the space of configurational probabilities for general classical spin models, provided that there is a symmetry associated with each spin degree of freedom. In some cases, we may exploit this symmetry even when the model Hamiltonian contains an external field contribution that enthalpically breaks the symmetry. 

For concreteness, we first discuss how the approach applies to a simple finite dipolar chain.

\subsection{\label{sec:dipole model} Illustrative model: finite dipolar chain}
Let us consider a reference system of interacting dipoles on a finite one-dimensional regular lattice, $\Lambda$, which we assume to extend in the $x$-Cartesian direction.
Each dipole has a fixed magnitude, $d_i$, and can rotate freely within a two-dimensional plane as sketched in Fig.~\ref{fig:intro}. Under the point-dipole approximation, the interaction between dipoles $i$ and $j$ is,
\begin{eqnarray}
    V_{ij} = \frac{d_i d_j \left[\hat{\mu}_i^{\top} \hat{\mu}_j - 3(\hat{\mu}_i^{\top} \hat{\boldsymbol{x}}) (\hat{\mu}_j^{\top} \hat{\boldsymbol{x}}) \right]}{r_{ij}^3},
\end{eqnarray}
where $\hat{\mu}_i = (\cos{\vartheta_i}, \sin{\vartheta_i})$ denotes the orientation of dipole $i$, $\hat{\boldsymbol{x}}$ indicates the unit vector separating dipoles $i$ and $j$, and $r_{ij}$ is the separation distance. The indices $1\leq i \leq n$ specify position along the lattice so that $r_{ij} = a \vert  i - j \vert$ for lattice constant of $a$.
The system Hamiltonian is,
\begin{eqnarray}
    \mathcal{H}( \underline{\vartheta}|d_{\Lambda}) = \frac{1}{2}\sum_{i\neq j}^{n} V_{ij} - \sum_{i=1}^{n} d_i \hat{\mu}_i^{\top} \mathbf{E}^\mathrm{ext}_i,
    \label{eq:Hamiltonian}
\end{eqnarray}
where $\underline{\vartheta}=(\vartheta_1, \cdots, \vartheta_n)$ specifies the angular configuration of dipoles and the second term describes the influence of external electric field, $\mathbf{E}^\mathrm{ext}_i$.

With fixed temperature, lattice size, and number of dipoles, the canonical partition function is given by,
\begin{eqnarray}
    \mathcal{Z} = \int \prod_{i=1}^n d\vartheta_i \exp\left[-\beta \mathcal{H}(\underline{\vartheta}|d_{\Lambda})\right] \equiv \exp\left[-\beta F\right], 
    \label{eq:Zeq}
\end{eqnarray}
where the integral spans all possible dipole configurations and $1/\beta = k_\mathrm{B} T$ is the Boltzmann constant times temperature.
This relationship defines a free energy $F$, yet the equilibrium measure $\rho_\mathrm{eq} (\underline{\vartheta}) = \exp\left[-\beta \mathcal{H}(\underline{\vartheta}) \right]/\mathcal{Z}$ is computationally intractable for all but the smallest system sizes due to the high-dimensional integral $\mathcal{Z}$.
Such intractability can be typically circumvented by either biased subsampling or the application of mean-field theory (MFT).

Biased subsampling takes advantage of the fact that in many cases $\mathcal{Z}$ is dominated by a small subset of low energy configurations.
Sampling these configurations, \textit{e.g.}, via Monte Carlo algorithms, typically enables asymptotic convergence of equilibrium properties, despite that these techniques are prone to frustration in systems with nonconvex free energy landscape.
Application of MFT, on the other hand, reduces computation by creating a model system with dramatically simplified configuration probabilities and a corresponding free energy that can be variationally related to that of the target system, \textit{i.e.}, $F_\mathrm{MF} \geq F$.
A major advantage of MFT is that the model system is often solvable through numerical or analytical methods.

\section{\label{sec:MFT} Mean Field Formalism}

In MFT, the fluctuating environment is modeled by a static mean field.
Because the neglected fluctuations are entropically favorable, the free energy of the mean field system represents a loose upper bound on that of the interacting system.
The trade-off in accuracy is that MFT significantly reduces the complexity and system size scaling associated with the computation of system properties.

\begin{figure}[t]
\centering
\includegraphics[width=.375\textwidth]{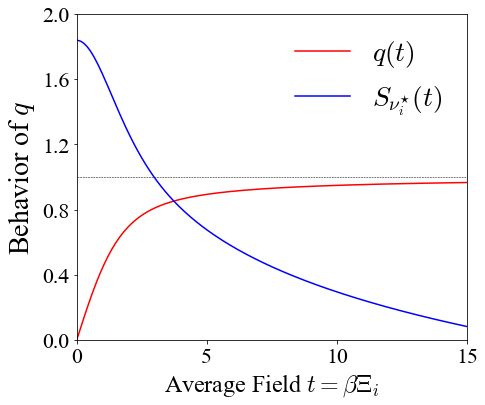}
\caption{Behavior of the function $q$ and the entropy $S_{\nu_i^{\star}}$ as a function of the nondimensional average field $\beta \Xi_i$ on tagged spin $i$. Observe that $q(t)$ is bounded between $0$ and $1$, where its derivatives further imply that $q(t)$ is monotonically increasing and concave. 
Consequently (combining Eqs.~\ref{eq:S_MF} and~\ref{eq:nu_star}), the MF entropy monotonically decreases as the averaged field increases. Dashed line marks the asymptotic behavior $q \rightarrow 1$. } 
\label{fig:q}
\end{figure}

For the lattice dipole system, the mean field free energy can be expressed parametrically as, 
\begin{eqnarray}
    \mathcal{F}_{\rm MF}(\underline{\theta},\underline{r}) = \mathcal{H}_{\rm MF}(\underline{\theta},\underline{r}) - T S_{\rm MF}(\underline{\theta}, \underline{r}), 
    \label{eq:F_MF}
\end{eqnarray}
where the vectors $\underline{\theta}$ and $\underline{r}$ contain the mean orientations, $\theta_i = \langle \vartheta_i \rangle$, and polarizations, $r_i = d_i \sqrt{\langle \cos{\vartheta_i} \rangle^2 + \langle \sin{\vartheta_i} \rangle^2} \leq d_{i}$. In Eq.~\ref{eq:F_MF}, the enthalpic term, $\mathcal{H}_{\rm MF} = \mathcal{H}(\underline{\theta} | \underline{r})$, is simply the average system energy. The entropic term, $S_{\rm MF},$ is a sum of single dipole entropies (reflecting the lack of dipole correlations), 
\begin{equation}
    S_{\rm MF} = \sum_{i=1}^n S_{\nu_i^{\star}} =  -k_\mathrm{B} \sum_{i=1}^{n} \int d\vartheta \nu_i^{\star}(\vartheta) \ln \nu_i^{\star} (\vartheta),
\label{eq:S_MF}
\end{equation}
where $S_{\nu_i}$ labels the entropy of dipole $i$ dictated by an angular distribution $\nu_i$.
The distribution $\nu_i^{\star}$ in Eq.~\ref{eq:S_MF} denotes the unique maximum entropy distribution commensurate with the mean $(\theta_i, r_i)$, derived as the extremal of the action,
\begin{equation}
\begin{split}
    \hspace{-0.2 cm} \mathcal{A}_i[\nu; \underline{\theta}, \underline{r}] = & - T S_{\nu_i} +  \lambda_+ \left[ 1 - \int d\vartheta \nu_i(\vartheta)\right] \\
    & + \lambda_- \left[ r_i \exp{(i\theta_i)} - d_i \int d\vartheta \nu_i(\vartheta) \exp{(i\vartheta)} \right],
    \label{eq:action}
\end{split}
\end{equation}
where the Lagrange multipliers $\lambda_{\pm}$ enforce the normalization and the first moment constraint. Specifically, the maximizing $\nu_i^{\star}$, also known as a \textit{von Mises} distribution, takes the form, 
\begin{equation}
    \nu_{i}^{\star}(\vartheta) = \frac{\exp{\left[ \beta \Xi_i \cos {(\vartheta - \theta_i)} \right]}}{2\pi I_0(\beta \Xi_i)},
    \label{eq:nu_star}
\end{equation}
where $\theta_i$ gives the angular mean as expected for consistency, $\Xi_i = d_i \olsi{E}_i$ sets the angular fluctuation under an averaged field $\olsi{E}_i \propto q^{-1}(r_i/d_i)$ for which $q(t)$ is a differentiable function satisfying the Riccati equation $\nabla_t q = 1- q/t - q^2$ with initial condition $q(0) = 0$, and $I_0$ in the normalizing constant denotes the zeroth order modified Bessel function of the first kind. The behavior of the function $q(t)$ is shown in Fig.~\ref{fig:q}.

\section{\label{sec:main result} Main Result and Applicability}
In essence, the mean field approximation limits the space of possible angular distributions to those satisfying Eq.~\ref{eq:nu_star} (discussed further in Appendix A).
This limitation significantly reduces the search space for the optimization of $\mathcal{F}_\mathrm{MF}$ in Eq.~\ref{eq:F_MF}.
However, even in this reduced search space, $\mathcal{F}_\mathrm{MF}$ may exhibit metastable minima that prevent typical optimization routines, such as gradient updates, from reaching the global minimum. Our goal here is to outline a robust procedure that, when properly initialized, will reliably converge to the optimum under the mean field approximation. We formulate the procedure in terms of a mean field iterator, $\mathcal{G}_\mathrm{MF}$, which generates directed walks in the MF state space, \textit{i.e.},
\begin{eqnarray}
    \mathcal{G}_{\rm MF}\left[ (\underline{\theta}, \underline{r})_{\tau} \right]=(\underline{\theta}, \underline{r})_{\tau+1},
    \label{eq:MF_it}
\end{eqnarray}
where the index $\tau$ labels the iteration step. In our case, we consider $\mathbf{E}^\mathrm{ext}_i = E_i \hat{\boldsymbol{x}}$ such that $E_i \geq 0$ on all lattice sites, such as would result from a voltage drop across the lattice driven by a pair of electrochemical reservoirs. This assumption will be kept implicit throughout the remaining sections.

Here, we provide a summary of our approach. Relevant key ideas are presented by order in the next sections, while technical details are described in their entirety within the Appendix.
Under our assumptions, the explicit form of the iterator, $\mathcal{G}_{\rm MF}$, in Eq.~\ref{eq:MF_it} is given by,
\smallskip
\begin{eqnarray}
    \begin{bmatrix}
    \mathcal{G}_{{\rm MF},1} \\
    \vdots \\
    \mathcal{G}_{{\rm MF},n}
    \end{bmatrix} = \begin{bmatrix}
    d_1 q(\beta d_1 E_{\rm MF,1}) \\
    \vdots \\
    d_n q(\beta d_n E_{\rm MF,n})
    \end{bmatrix},
\end{eqnarray}
which guides directed walks in MF state space. For the dipolar model, $q$ is the function appearing within Eq.~\ref{eq:nu_star} and illustrated in Fig.~\ref{fig:q}, whereas
\begin{eqnarray}
    E_{{\rm MF},i}(\underline{\theta}, \underline{r}) =   E_i + 2 \sum_{j \neq i} \frac{r_j \cos{\theta_j}}{r_{ij}^3},
\end{eqnarray}
represents the strength of molecular mean field experienced by dipole $i$. We will show that the mean fields all point in the $\hat{x}$ direction and depend only on the $x$-components of the dipole averages. The iterator follows from minimization of the free energy $\mathcal{F}_{\rm MF}$ and is constructed step by step in Secs.~\ref{subsec:optimizer}-\ref{subsec:iterator}, while its convergence is established in Sec.~\ref{subsec:theorems}. 

The results for dipolar chain extend to other classes of spin models for which spin-spin interaction orients favorably along direction of the imposed external field (ferromagnetic models with consistent external field are thus canonical examples), as is elaborated in Sec.~\ref{subsec:genearlized lattice models}. Our lattice description then motivates a discussion on the continuum fields in Sec.~\ref{subsec:continuum fields}.

\section{\label{sec:iterator} Iterator and Its Convergence}
\subsection{\label{subsec:optimizer} Global optimizer}

Certain symmetries that are inherent to dipolar (spin-spin) interactions result in a significant reduction of the search space in the optimization of $\mathcal{F}_{\rm MF}$.
Our construction of the MF iterator, $\mathcal{G}_{\rm MF}$, exploits these symmetries, which are formalized as the lemma below.

\begin{figure}[H]
\centering
\includegraphics[width=.475\textwidth]{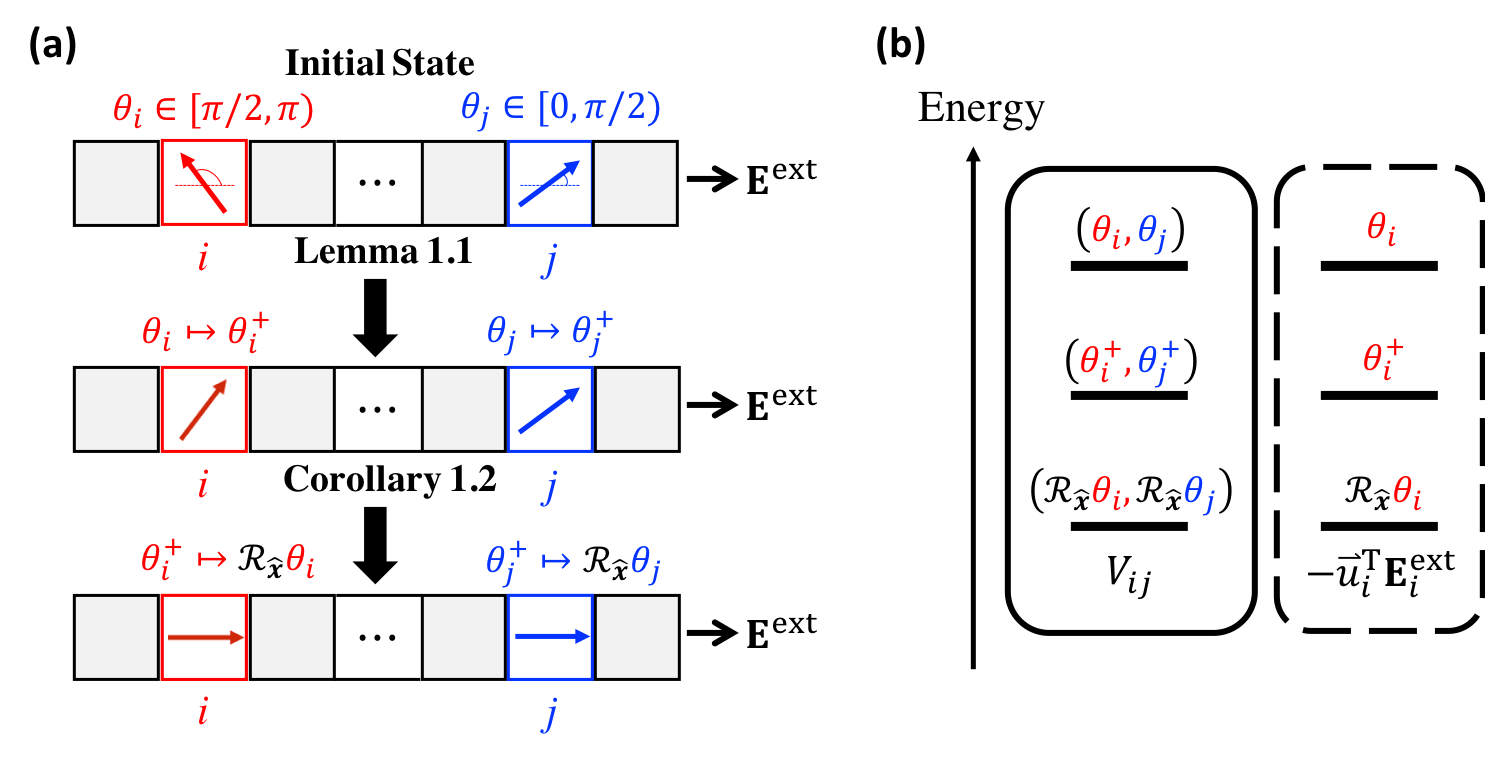}
\caption{Elementary state space operations that relax the mean-field free energy of finite dipolar chains.
$(a)$ Schematic illustrations of $\hat{\boldsymbol{y}}$-reflection and $\hat{\boldsymbol{x}}$-rotation acting on the mean states $(\theta_i, \theta_j)$ of a pair of tagged dipoles $i$ and $j$.
$(b)$ Energy diagrams for the two-body dipolar contribution and one-body external field contribution upon the elementary operations.}
\label{fig:Lemma1_Sketch}
\end{figure}

\textbf{Lemma 1.1.} A global maximizer $(\underline{\theta}^{\ast}, \underline{r}^{\ast})$ of
\begin{eqnarray}
    \Phi_{\rm MF}(\underline{\theta}, \underline{r}) = - \beta \mathcal{F}_{\rm MF}(\underline{\theta}, \underline{r}),
    \label{eq:MF_optimization}
\end{eqnarray}
satisfies $0 \leq r_i^{\ast}\cos{\theta_i^\ast} \leq d_i$ for all $1 \leq i \leq n$. 

\noindent \textit{Sketch of proof.} Here we convey the central idea of the proof. 
An elaborated proof can be found in Appendix B. 

We first recognize the invariance of the entropies $S_{\nu_i^{\star}}$ under the partial reflections,
\begin{eqnarray}
    \hspace{-0.5cm} (\cos{\theta_i}, \sin{\theta_i}) \mapsto \begin{cases}
    (-\cos{\theta_i}, \sin{\theta_i}) &  {\rm if}~ \cos{\theta_i} \leq 0 \\
    (\cos{\theta_i}, \sin{\theta_i}) & {\rm otherwise}
    \end{cases},
\end{eqnarray}
which can be realized through circular shift of the angular distributions and diagrammatically understood in Fig.~\ref{fig:Lemma1_Sketch}(a). For a mean configuration $\underline{\theta}$, we consider such reflections on all sites and denote the partially reflected mean configuration by $\underline{\theta}^{+}$.

\begin{figure}[H]
\centering
\includegraphics[width=.35\textwidth]{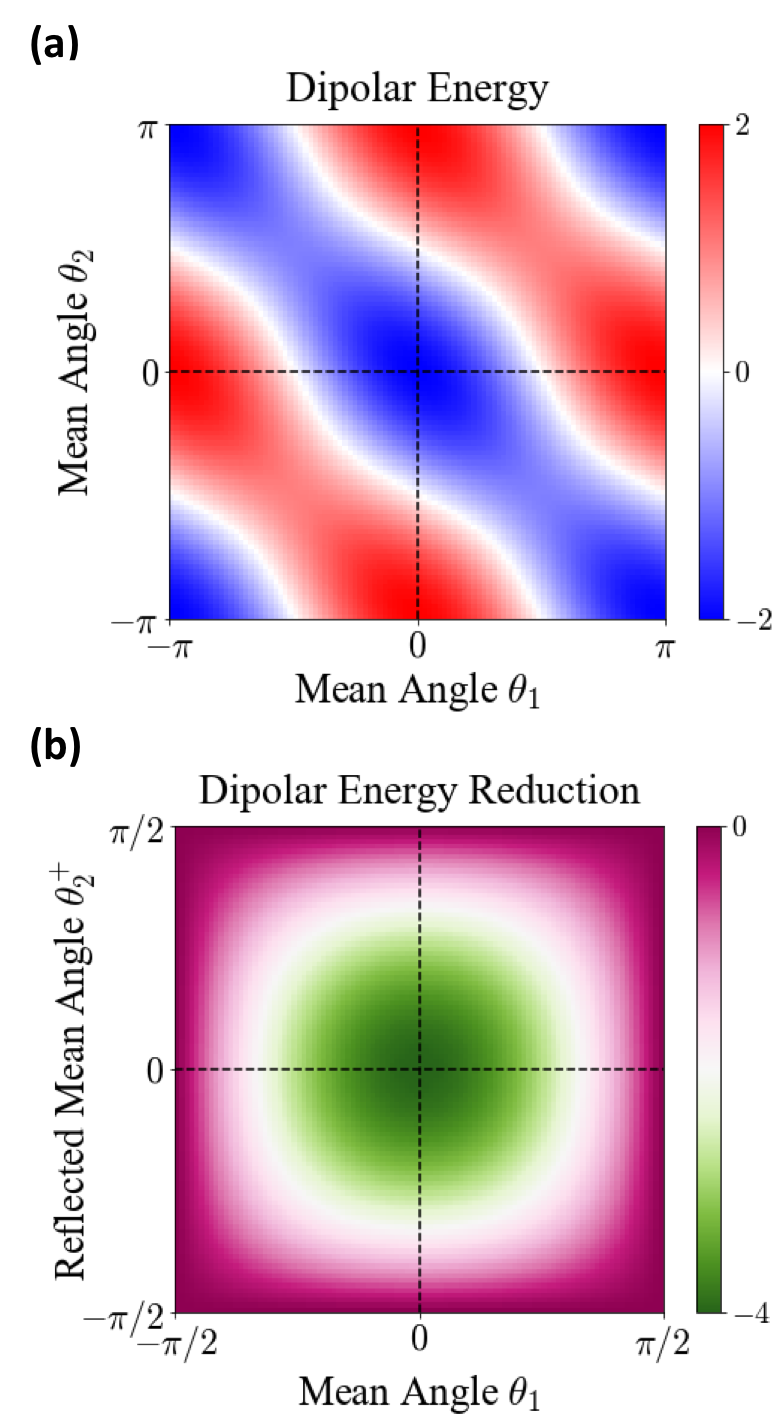}
\caption{Symmetry of the dipolar interaction. $(a)$ Bare potential $V_{12}(\theta_1, \theta_2)$ plotted over the orientations $(\theta_1, \theta_2)$ of two mean dipoles separated by a unit distance. $(b)$ Potential reduction $V_{12}(\theta_1, \theta_2^{+}) - V_{12}(\theta_1, \theta_2)$ upon reflecting one dipole orientation $\theta_2 \mapsto \theta_2^{+}$ while fixing the other dipole orientation $-\pi/2 \leq \theta_1 \leq \pi/2$.}
\label{fig:bond}
\end{figure}

The operation $\underline{\theta} \mapsto \underline{\theta}^+$ leads to an energy reduction since it induces favorable dipolar couplings $V_{ij}$ and encourages dipole alignment with the external field $\mathbf{E}^\mathrm{ext}_i$ as depicted in Fig.~\ref{fig:Lemma1_Sketch}(b). 
In Fig.~\ref{fig:bond}, we show the $V_{ij}$ reduction by plotting the full range of coupling values for two dipoles (Fig.~\ref{fig:bond}$a$) and the change in coupling from single dipole reflection (Fig.~\ref{fig:bond}$b$). Therefore, $\mathcal{F}_{\rm MF}(\underline{\theta}^{+}, \underline{r}) \leq \mathcal{F}_{\rm MF}(\underline{\theta}, \underline{r})$ and $0 \leq r_i^{\ast}\cos{\theta_i^\ast} \leq d_i$. $\square$

\bigskip

In fact, the global maximum is attained only when the mean dipoles align with the external field.

\textbf{Corollary 1.2.} $\theta_i^{\ast} = 0$ for all $1 \leq i \leq n$. 

\noindent \textit{Sketch of proof}. A mean configuration in full alignment results from rotations onto the vector $\hat{\boldsymbol{x}}$, 
\begin{eqnarray}
    (\cos{\theta_i}, \sin{\theta_i}) \mapsto (1, 0),
\end{eqnarray}
denoted as $\underline{\theta} \mapsto \mathcal{R}_{\hat{\boldsymbol{x}}}\underline{\theta}$ on the lattice. Invariance of entropies and reduction of energetics follow from these rotations, with a diagrammatic view presented in Figs.~\ref{fig:Lemma1_Sketch} and \ref{fig:bond}. $\square$

\subsection{\label{subsec:iterator} Mean field iterator}
The MF iterator $\mathcal{G}_\mathrm{MF}$ is a state-space operator that generates discrete flow towards the optima of $\Phi_{\rm MF}$. We use results in the previous section to derive its form. Although the parametrized distributions $\nu_{i}^{\star}$ and free energy function $\mathcal{F}_{\rm MF}$ are most naturally expressed in the polar coordinate, it is advantageous to consider a Cartesian frame where one of the axes points in the external field direction $\hat{\boldsymbol{x}}$. 
With this coordinate change, 
\begin{eqnarray}
    (X_i, Y_i)= r_i(\cos{\theta_i},
    \sin{\theta_i}),
    \label{eq:Jacobian}
\end{eqnarray}
the projections $X_i$ can be isolated and treated separately. Now taking $\nabla \Phi_{\rm MF} = 0$ for the MF optimization in Eq.~\ref{eq:F_MF}, we arrive at an array of self-consistent equations that manifest the first-order condition $\beta \nabla \mathcal{H}_{\rm MF} = \beta T \nabla S_{\rm MF}$,
\begin{eqnarray}
    \displaystyle \beta \sum_{j=1}^{n} 
    \doubleunderline{T}^{i,j}  \begin{bmatrix}
    X_j \\
    Y_j
    \end{bmatrix} = \begin{bmatrix}
     \beta E_i  + \nabla_{X_i} S_i \\
     \nabla_{Y_i} S_i
    \end{bmatrix},
    \label{eq:SCE}
\end{eqnarray}
where the $2 \times 2$ matrix $\doubleunderline{T}^{i,j}$ encodes the dipolar coupling,
\begin{eqnarray}
    \doubleunderline{T}^{i,j} = \lim_{\epsilon \rightarrow 0^+} \displaystyle \frac{1-\delta_{ij}}{\epsilon + r_{ij}^3} \begin{bmatrix}
    -2 & 0 \\
     0 & 1 \\ 
    \end{bmatrix},
\end{eqnarray}
with $\delta_{ij}$ denoting the Kronecker delta, $E_i$ accounts for the local external fields, and $S_i = S_{\nu_i^{\star}}$ gives the single dipole entropies,
\begin{eqnarray}
    S_i = -  K_i q^{-1}(K_i) + \ln{ \frac{I_0 ( q^{-1}(K_i) )}{2\pi}},
\end{eqnarray}
with nondimensional polarity $K_i = \sqrt{X_i^2+ Y_i^2}/d_i$ characterizing the angular dispersion. Let $\underline{\Gamma} = (X_1,Y_1,\cdots,X_n,Y_n)$ be the paired coordinates in shorthand so Eq.~\ref{eq:SCE} can be arranged as a matrix equation,
\begin{eqnarray}
    \beta \doubleunderline{\Tilde{T}} \cdot \underline{\Gamma}  = \beta \Tilde{E} + \Tilde{C}(\underline{\Gamma}), 
    \label{eq:SCE2}
\end{eqnarray}
where the $2n \times 2n$ matrix $\doubleunderline{\Tilde{T}}$ composed of $2 \times 2$ blocks,
\begin{eqnarray}
    \doubleunderline{\Tilde{T}} = \begin{bmatrix}
    0 & \doubleunderline{T}^{1,2} & \cdots & \doubleunderline{T}^{1,n}  \\
    \doubleunderline{T}^{2,1} & 0 &  & \vdots \\
    \vdots & & \ddots & \doubleunderline{T}^{n-1,n} \\
    \doubleunderline{T}^{n,1} & \cdots & \doubleunderline{T}^{n, n-1} & 0
    \end{bmatrix},
\end{eqnarray}
accommodates the collective anisotropic interaction in the dipolar chain, $\Tilde{E} = (E_1,0,\cdots,E_n,0)$ contains the augmented external fields, and
\begin{eqnarray}
    \Tilde{C} = (\nabla_{X_1}S_1, \nabla_{Y_1}S_1, \cdots,\nabla_{X_n}S_n, \nabla_{Y_n}S_n),
\end{eqnarray}
designates the entropic forces.

Eq.~\ref{eq:SCE2} defines the state-space property of the MF solution, $\underline{\Gamma}^*$, and can be utilized to ensure that $\underline{\Gamma}^*$ is a fixed point of $\mathcal{G}_\mathrm{MF}$.
Ideally, one would convert Eq.~\ref{eq:SCE2} into the form $\underline{\Gamma} = \mathcal{G}_{\rm MF}(\underline{\Gamma})$ and analyze the associated fixed-point iteration. 
However, due to the noninvertibility of the partitioned matrix $\doubleunderline{\Tilde{T}}$ and entropy gradient $\Tilde{C}$, instead, we refer to Corollary 1.2 in Sec.\ref{subsec:optimizer} and consider the dimensionally reduced $\mathcal{G}_{\rm MF}$ through a projection of Eq.~\ref{eq:SCE} onto the ``important" subspace $Y_i \equiv 0$,
\begin{eqnarray}
    \beta E_{{\rm MF}, i}(\underline{X}) = \beta E_i + 2 \beta \sum_{j \neq i} \frac{X_{j}}{r_{ij}^3} = \frac{q^{-1}(X_i/d_i)}{d_i}, 
    \label{eq:Eff_Efield}
\end{eqnarray}
where the molecular field $E_{{\rm MF},i}$ acting on dipole $i$ can be explicitly defined through Eq.~\ref{eq:Eff_Efield} after we project out half of the equations satisfied vacuously, \textit{i.e.}, $\beta \sum_{j \neq i} Y_{j}/r_{ij}^{3} = 0$. Hence we obtain our expression of the MF iterator,
\begin{eqnarray}
    \mathcal{G}_{{\rm MF}, i} = d_i q\left( \beta d_i E_{{\rm MF},i} \right),
\end{eqnarray}
where $\mathcal{G}_{{\rm MF}, i} \leq d_i$ since $q \leq 1$. We then define $\mathcal{G}_{\rm MF}(\underline{X}) = (\mathcal{G}_{{\rm MF}, 1}(\underline{X}), \cdots, \mathcal{G}_{{\rm MF}, n}(\underline{X}))$ in a component-wise manner. Clearly the properties of $\mathcal{G}_{\rm MF}$ depend on the dipolar model parameters $(\beta, E, d_{\Lambda},a)$, and again we assume $E_{i} \geq 0$ throughout the rest of the work. For practicality, we proceed to establish theoretical bounds on the convergence of $\mathcal{G}_{\rm MF}$ by resorting to the approximation theorems below.

\subsection{\label{subsec:theorems} Convergence theorems}

Our first theorem states that the iteration generated by $\mathcal{G}_{\rm MF}$ converges uniformly under strong external field, regardless of the initial condition.
\bigskip

\textbf{Theorem 2.1.} There is an external field $E^{\ast} = (E^{\ast}_1,\cdots,E^{\ast}_n)$ such that if $E \geq E^*$ entry-wise, the MF iterator,
\begin{eqnarray}
    \underline{X}_{\tau + 1} = \mathcal{G}_{\rm MF}(\underline{X}_{\tau}),
    \label{eq:discrete_MF_flow}
\end{eqnarray}
with any initial estimate $\underline{X}_0$ will converge linearly to the MF solution $\underline{X}^{\ast} = \mathcal{G}_{\rm MF}(\underline{X}^{\ast})$ as the unique fixed point. That is,
\begin{eqnarray}
    {\| \mathcal{G}_{\rm MF}^{(\tau)}(\underline{X}_{0}) - \underline{X}^{\ast} \|}_{p} \leq \mathcal{B}_\tau \|{\mathcal{G}_{\rm MF}(\underline{X}_{0}) - \underline{X}_0 \|}_{p},
    \label{eq:Thm2.1}
\end{eqnarray}
where
\begin{eqnarray}
    \norm{\underline{X}}_{p} = \left[ \sum_{i=1}^n |X_i|^p \right]^{1/p},
\end{eqnarray}
gives the vector $\ell^p$-norm for integer $1 \leq p \leq \infty$, $\mathcal{G}_{\rm MF}^{(\tau)}$ denotes the $\tau{\rm th}$ repeated application of $\mathcal{G}_{\rm MF}$, and $\mathcal{B}_{\tau}$ is a $\tau$-dependent bound controlling how rapidly Eq.~\ref{eq:discrete_MF_flow} converges towards the MF solution, up to a constant set by the initial condition $\underline{X}_0$ as written on the RHS of Eq.~\ref{eq:Thm2.1}. $\mathcal{B}_{\tau}$ decreases with $\tau$, reaching zero as $\tau \rightarrow \infty$, and is sensitive to dipolar model parameters. When $\mathcal{B} = 2n \max_{i} d_i / E^{\ast}_i < 1$ and $a = 1$, a possible choice is $ \mathcal{B}_{\tau} = \mathcal{B}^{\tau}/(1-\mathcal{B})$.

\begin{figure}[H]
\centering
\includegraphics[width=.365\textwidth]{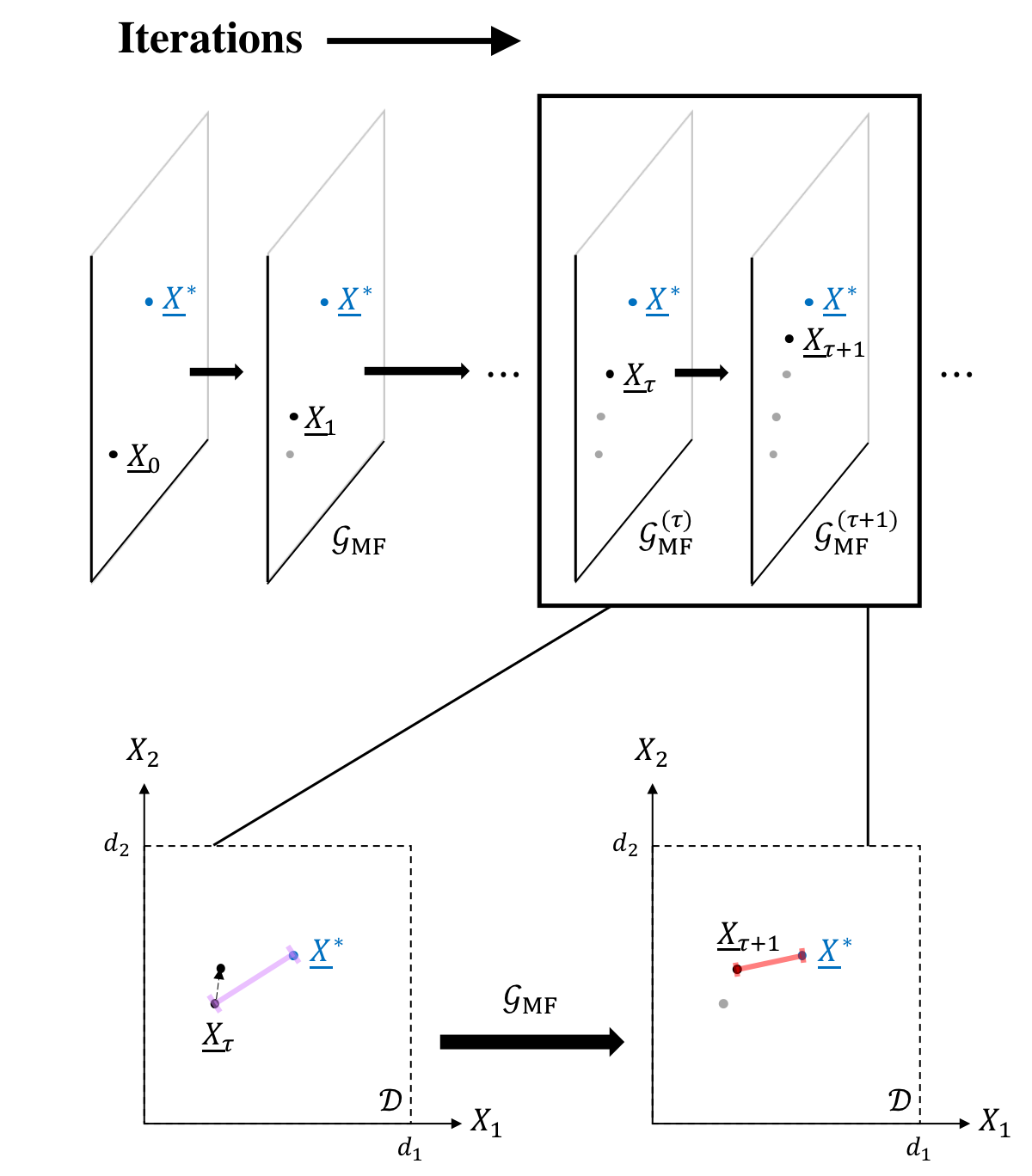}
\caption{Illustrating Theorem 2.1 with a representative directed walk $(\underline{X}_{\tau})_{\tau}$ via the MF iterator.
Updated state marked in black approaches the equilibrium state $\underline{X}^{\ast}$ marked in blue under the iterations. 
Contractivity of $\mathcal{G}_{\rm MF}$ is illustrated in the lower inset, where the $\ell^2$-distance to $\underline{X}^{\ast}$ is marked with line segments. From left to right, segment color varies from purple (longer distance) to red (shorter distance).
}
\label{fig:Thm1_Sketch}
\end{figure}

\noindent \textit{Sketch of proof}. When $\mathbf{E}_i^{\rm ext}$ is sufficiently large, we expect $F \approx F_{\rm MF}$ due to dielectric saturation. We prove the theorem by showing that $\mathcal{G}_{\rm MF}$  is contractive, namely any pair of states gets mapped closer to each other under $\mathcal{G}_{\rm MF}$ so eventually $\underline{X}_{\tau} \rightarrow \underline{X}^{\ast} = \mathcal{G}_{\rm MF}(\underline{X}^{\ast})$ as illustrated in Fig.~\ref{fig:Thm1_Sketch}. The update $\underline{X}_0 \mapsto \underline{X}_1 \mapsto \cdots \mapsto \underline{X}_{\tau} \mapsto \cdots$ in this regime is monotonic as $\underline{X}_{\tau}$ improves after each additional iteration (See Appendix B). $\square$

\bigskip

Now we examine the convergence when the external field is weak and does not exhibit any scaling with the system size. In particular, we follow Koehler's approach~\cite{Koehler2019} by exploiting curvature of the MF free energy surface and deliberately picking out a subdomain of initial estimates. The following lemma counts the number of fixed  points in the MF state space, which helps eliminate ambiguity when we specify suitable initial estimates in our next theorem.

\bigskip

\textbf{Lemma 2.2.}  $\mathcal{G}_{\rm MF}(\underline{X})$ yields either zero or one fixed point in the interior of its domain.

\noindent \textit{Sketch of proof}. Suppose that we find two fixed points, $\underline{X}_c$ and $\underline{X}_c'$. We then take an interpolating path $\underline{\rchi}_{0 \leq \lambda \leq 1} = \lambda \underline{X}_c + (1-\lambda) \underline{X}_c'$ for which we assume $(\underline{X}_c')_{i} < (\underline{X}_{c})_{i}$ for some site $i$ and may suitably extend the function,
\begin{eqnarray}
    I(\lambda) = \mathcal{G}_{{\rm MF}, i}(\underline{\rchi}_{\lambda}) - (\underline{\rchi}_{\lambda})_{i},
\end{eqnarray}
outside $[0,1]$. We may locate a $\lambda_0 \in (0,1)$ so that $\nabla_{\lambda} I(\lambda_0) = 0$ by the mean value theorem. However, we also show concavity of $I$ along the path $\underline{\rchi}_{\lambda}$, implying a new constraint $\nabla_{\lambda} I(0) < 0$ contradicting the internal constraint $\nabla_{\lambda} I(\lambda_0) = 0$. $\square$

\begin{figure}[H]
\centering
\includegraphics[width=.365\textwidth]{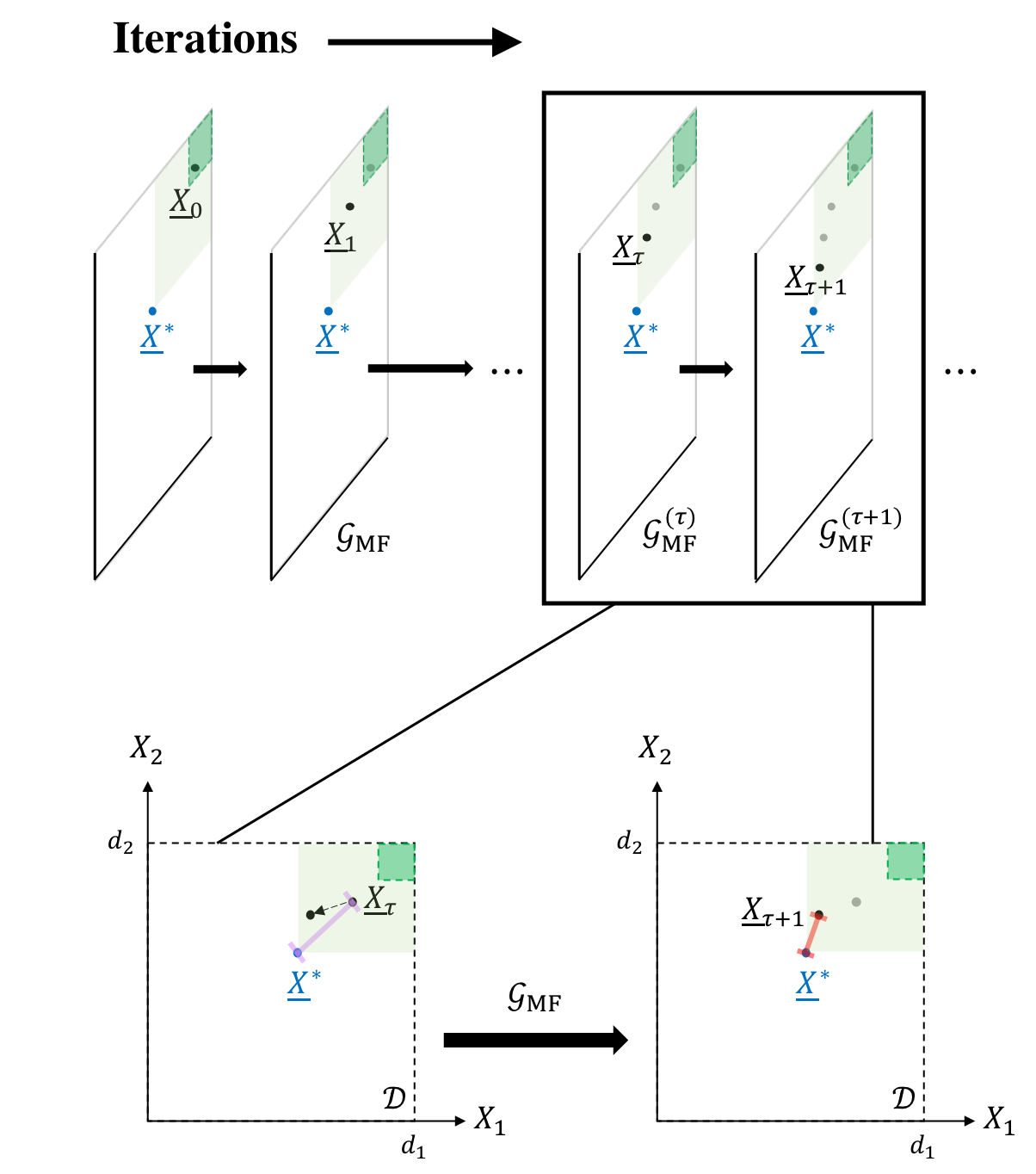}
\caption{Illustrating Theorem 2.3 with a representative directed walk $(\underline{X}_{\tau})_{\tau}$ via the MF iterator. 
Updated state marked in black diagonally approaches the state $\underline{X}^{\ast}$ marked in blue. The stable region $\Delta_{\mathcal{D}}$ is shaded in pale green and its stability is illustrated in the lower inset, where the $\ell^2$-distance to $\underline{X}^{\ast}$ is marked with line segments for segment color varying from purple (longer distance) to red (shorter distance). The subregion of initial estimates is shaded in bright green.
}
\label{fig:Thm2_Sketch}
\end{figure}

Given $E_i > 0$, the lemma immediately implies the existence and uniqueness of an interior critical point, and we will make reference to this point $\underline{X}^{\ast}$ for the prescription of suitable initial estimates in the next theorem.

\bigskip

\textbf{Theorem 2.3.}  There exists a region $\Delta_{\mathcal{D}}$ that is stable under application of the MF iterator,
\begin{eqnarray}
    \underline{X}_{\tau + 1} = \mathcal{G}_{\rm MF}(\underline{X}_{\tau}),
    \label{eq:discrete_MF_flow2}
\end{eqnarray}
so a suitable initial estimate $\underline{X}_0$ in $\Delta_{\mathcal{D}}$ converges to the MF solution $\underline{X}^{\ast} = \mathcal{G}_{\rm MF}(\underline{X}^{\ast})$ on the level of free energy. That is, $\underline{X}_{0} \in \Delta_{\mathcal{D}}$ implies $\mathcal{G}_{\rm MF}(\underline{X}_{\tau}) \in \Delta_{\mathcal{D}}$ for $\tau \geq 0$ and
\begin{eqnarray}
     \abs{\Phi_{\rm MF}( \mathcal{G}^{(\tau)}_{\rm MF}(\underline{X}_{0})) - \Phi_{\rm MF}(\underline{X}^{\ast})} \leq \frac{\Tilde{\mathcal{B}}}{\tau},
     \label{eq:Thm2.3}
\end{eqnarray}
where $\Tilde{\mathcal{B}}$ is a $\tau$-independent bound that controls how rapidly the free energy along the discrete state space flow from Eq.~\ref{eq:discrete_MF_flow2} converges to the true MF free energy. When $a = 1$, a possible choice is $\Tilde{\mathcal{B}} = n \beta \zeta(3) \max_{i} d_i^2$ where
\begin{eqnarray}
    \zeta(s) = \sum_{k=1}^{\infty} \frac{1}{k^s},
\end{eqnarray}
denotes the Riemann zeta function.

\noindent \textit{Sketch of proof}. In the absence of strong external fields, the MF update is not necessarily monotonic. This motivates the preference for a convex subdomain $\Delta_{\mathcal{D}} = \{\underline{X}: \underline{X} \geq \underline{X}^{\ast} \}$ where monotonicity is preserved ($``\geq"$ understood entry-wise). The stability of $\Delta_{\mathcal{D}}$ follows from the fact $\nabla_t q(t) > 0$ for $t \geq 0$, as a majorizing relation of the mean field strengths, 
\begin{eqnarray}
    \underline{X}_{\tau} \geq \underline{X}^{\ast} \implies E_{{\rm MF}}(\underline{X}_{\tau}) \geq E_{{\rm MF}}(\underline{X}^{\ast}),
\end{eqnarray}
induces that of the site responses, $\mathcal{G}_{\rm MF}(\underline{X}_{\tau}) = \underline{X}_{\tau+1} \geq \underline{X}^{\ast} = \mathcal{G}_{\rm MF}(\underline{X}^{\ast})$, shown in Fig.~\ref{fig:Thm2_Sketch}.  The remaining of the proof follows from the concavity of MF free energy surface over the region $\Delta_{\mathcal{D}}$, which allows the safe marching of our suitably initialized flow $\underline{X}_{\tau}$ towards the optimal $\underline{X}^{\ast}$ (see Appendix B). $\square$

\section{\label{sec:genearlize} Beyond Dipolar Chains}
\subsection{\label{subsec:genearlized lattice models} Generalized lattice models}

An immediate attempt to extend validity of the convergence criteria is to address models with higher dimensional spins. In particular, we take a positive integer $p$, especially $p \geq 3$, and consider the corresponding angular degrees of freedom $\underline{\omega}$ on a $(p-1)$-sphere under the Hamiltonian, 
\begin{equation}
    \mathcal{H}(\underline{\omega})
    = \sum_{(i,j) \in \mathcal{E}} d_i d_j \hat{\mu}_i^{\top}  \mathcal{T}_{ij} \hat{\mu}_j - \sum_{i \in \mathcal{V}} d_i \hat{\mu}_i^{\top} \mathbf{E}_i^{\rm ext},
\end{equation}
where $\hat{\mu}_i(\omega_i)$ represents a $p$-dimensional unit vector on the  $(p-1)$-sphere and $\Lambda = (\mathcal{V}, \mathcal{E})$ is some undirected graph with vertices $\mathcal{V}$ and edges $\mathcal{E}$. Here we make an additional assumption that $\mathcal{T}_{ij} \in \mathbb{R}^{p \times p}$ discloses some orientational preference of the spin-spin interaction along the bond direction $\vec{r}_{ij}$. This includes the class of common vector models, \textit{e.g.}, the $\mathbb{Z}_2$-Ising/XY/Heisenberg model ($p=1,2,3$ respectively) with a choice of $\mathcal{T}_{ij} \equiv -J_{ij} \cdot I_{p \times p}$ for $J_{ij} \geq 0$. Following notational conventions in Sec.\ref{subsec:iterator}, we find that the iterator $\mathcal{G}_{{\rm MF}}$ derived from ensuring optimality of the free energy function assumes an identical form,
\begin{eqnarray}
    \mathcal{G}_{{{\rm MF},i}}(\underline{X}) = d_i \mathcal{Q}_{p}(\beta d_i E_{{\rm MF},i}),
    \label{eq:General_MF}
\end{eqnarray}
where $\underline{X}$ is the projected coordinate determined by the mean (hyper)spherical angles $(\phi_{i|1}, \cdots, \phi_{i|p-2}, \theta_i)$ and magnitude $r_i$,
\begin{eqnarray}
    X_{i} = r_i \cos{\theta_i} \prod_{I=1}^{p-2} \sin{\phi_{i|I}},
\end{eqnarray}
$\displaystyle E_{{\rm MF},i}(\underline{X})$ denotes the effective local field, and
\begin{eqnarray}
    \mathcal{Q}_{p}(t) = \frac{I_{p/2 }(t)}{I_{p/2 - 1}(t)},
\end{eqnarray}
is the activation function that renormalizes the mean local response with $I_{p}$ denoting the modified Bessel function of order $p$. These nonlinear functions $\mathcal{Q}_{p}$ share the key properties that $(i)$ $ 0 \leq \mathcal{Q}_{p}(t) \leq 1$, $(ii)$ $\nabla_t \mathcal{Q}_{p}(t) > 0$, and $(iii)$ $\nabla_t \nabla_t \mathcal{Q}_{p}(t)< 0$ for $t >0$. In the cases $p \leq 3$, we retrieve familiar functions,
\begin{eqnarray}
    \mathcal{Q}_{p}(t) = \begin{cases}
    B_{1/2}(t) \equiv \tanh{t};&~p=1 \\
    q(t);&~p=2 \\
    B_{\infty}(t) \equiv L(t);&~p=3
    \end{cases},
\end{eqnarray}
where $B_{1/2}(t)$ denotes the Brillouin function of order $1/2$ and $L(t) = \coth{t}-1/t$ denotes the Langevin function. We note that an iterator of the form of Eq.~\ref{eq:General_MF} also handles the class of discrete models for which the single spin space is a finite subset of the $(p-1)$-sphere. This includes the $N$-state Potts model ($p = 2$), with $\vartheta_{i} \in 2\pi \mathbb{Z}_N/N$, and its higher dimensional analogs. In a discrete case, the corresponding MF activation function $\mathcal{Q}_{\hat{p}}(t)$ satisfies properties $(i)$-$(iii)$ certainly when discretization on the sphere is spatially symmetric, \textit{e.g.},
\begin{eqnarray}
    \mathcal{Q}_{\hat{2}}(t) = \frac{\sum_{\vartheta \in 2\pi \mathbb{Z}_N/N} \cos{\vartheta} \exp{\left[ t \cos{\vartheta} \right] }}{\sum_{\vartheta \in 2\pi \mathbb{Z}_N/N} \exp{\left[ t \cos{\vartheta} \right]} },
\end{eqnarray}
where $\mathcal{Q}_{p}(t) \leq \mathcal{Q}_{\hat{p}}(t) \leq \tanh(t)$. 

Of course it is worth checking whether the convergence results established in the previous sections generalize to more abstract and complicated phase spaces, which we denote by $\mathcal{X}$. Suppose that $\mathcal{X}= V_{k}(\mathbb{R}^{p}) = \{ \mathcal{W} \in \mathbb{R}^{p \times k} : \mathcal{W}^{\top} \mathcal{W} = I_{k \times k} \}$ is a Stiefel manifold, \textit{i.e.}, the set of orthonormal $k$-frames in $\mathbb{R}^{p}$ that reduces to a sphere when $k = 1$. Spins valued on special Stiefel manifolds constitute a basic ingredient of the minimal models describing frustrated systems, where the ground state exhibits order in a non-planar way~\cite{TetraMagnetic,NonplanarMagnetic}. For example, consider the Hamiltonian,
\begin{eqnarray}
    \hspace{-0.55 cm} \mathcal{H}(\underline{\mathcal{W}}) = -\sum_{(i,j) \in \mathcal{E}} d_i d_j {\rm tr}\big( \mathcal{W}_i^{\top} \mathcal{W}_j \big) - \sum_{i \in \mathcal{V}} d_i {\rm tr}\big( E_i^{\top} \mathcal{W}_i \big),
    \label{eq:tetrahedral}
\end{eqnarray}
where $\mathcal{W}_i \in V_{3}(\mathbb{R}^3)$ represents the local frame of a tetrahedron spin of chirality ${\rm \det}(\mathcal{W}_i)$ on site $i$, $E_i \in \mathbb{R}^{3 \times 3}$ is a matrix giving the axial-specific external field. In MFT, the derived one-body distribution takes the parametrized form~\cite{mardiadirectional},
\begin{eqnarray}
    \nu_i^{\star}(\mathcal{W}_i) = \frac{\exp\big[ \beta {\rm tr} \big( \Pi_i^{\top} \mathcal{W}_i \big) \big]}{_{0}F_1\big( 3/2; \beta^2 \Pi_i^{\top} \Pi_i /4 \big)},
    \label{eq:Matrix_Fisher}
\end{eqnarray}
where $_{0}F_{1}$ is the hypergeometric function
of matrix argument and the matrix parameter $\Pi_i \in \mathbb{R}^{3 \times 3}$ can be completely expressed in terms of the mean spin orientation $W_i = \langle \mathcal{W}_i \rangle_{\nu_i^{\star}}$. Assuming isotropy of the external field such that $E_i \mapsto E_i I_{3 \times 3}$, a global maximizer $\underline{W}^{\ast}$ of the MF free energy function can be shown positive semidefinite and in fact strictly diagonal with descending entries on the main diagonal (see Appendix D). After introducing the projected MF coordinates,
\begin{eqnarray}
    \underline{X} = (X_i)_{i \in \mathcal{V}} = \begin{split}
        \big(d_1(W_1)_{11}, ~&d_1(W_1)_{22}, d_1(W_1)_{33}, \\
        &\cdots \\
        d_n(W_n)_{11}, ~&d_n(W_n)_{22}, d_n(W_n)_{33} \big)
    \end{split},
\end{eqnarray}
we arrive at a vectorial MF iterator $\mathcal{G}_{{{\rm MF},i}} = d_i \mathcal{Q}(\beta d_i E_{{\rm MF},i})$ for which
\begin{eqnarray}
    \hspace{-0.25cm} E_{{\rm MF},i}(\underline{X}) =  E_i I_{3 \times 3} +  \sum_{j: (i,j) \in \mathcal{E}} \begin{bmatrix} 
    (X_j)_{1} & 0 & 0 \\
    0 & (X_j)_{2} & 0 \\
    0 & 0 & (X_j)_{3}
    \end{bmatrix},
\end{eqnarray}
and for a diagonal matrix $W$,
\begin{eqnarray}
    ~~\mathcal{Q}(W) = \left[- \nabla_{\underline{\kappa}} S_i \right]^{-1}(W_{11}, W_{22}, W_{33}),
\end{eqnarray}
where the one-body entropy $S_i$ depends on $p$ singular values $\underline{\kappa} = (\kappa_1, \cdots, \kappa_p)$ of the mean orientation (here $p = 3$). 
Each singular value controls the width of the distribution along a principal direction, with a larger value indicating higher concentration in the responsible direction.
Notice that $\mathcal{G}_{\rm MF}$ adopts the previous form of Eq.~\ref{eq:General_MF} for tetrahedron spins with frozen chirality, \textit{i.e.}, $\mathcal{W}_i \in V_2(\mathbb{R}^{3}) \cong SO(3)$ (discussed in Appendix D). On the other hand, both Theorem 2.1 and Theorem 2.3 hold if we replace the spin pace $SO(3)$ with $SU(2)$ and the matrix transpose with hermitian conjugate, since the diffeomorphism of $SU(2)$ to the $3$-sphere brings us back to the spherical scenario of Eq.~\ref{eq:General_MF}.

We have thus highlighted the compatibility of the iterative mean field approach with a wider class of spin models in statistical mechanics and solid state physics. Ideally we may ask the same question about other spin spaces. 
However, without the assured existence of a canonically invariant measure, \textit{e.g.}, a Lebesgue measure on $\mathbb{R}^{p}$, the validity of our approach is not guaranteed. As a consequence, we need stronger arguments to establish the invariance of the MF entropic term when spins are valued on a general compact Riemannian manifold.

\subsection{\label{subsec:continuum fields} From lattice models to continuum fields}
The lattice description above admits a natural field theory extension. For simplicity, we consider a scalar Euclidean field $f$ valued in the space of tempered distributions on $\mathbb{R}^d$. Recall that a tempered distribution $f$ has a canonical pairing with a Schwartz function $\mathcal{O}$ through, 
 \begin{equation}
     f[\mathcal{O}] = \int_{\mathbb{R}^d} d\xi f(\xi) \mathcal{O}(\xi).
 \end{equation}
In the familiar context of liquid state theory, $f$ can be thought of as the density profile of some electrolyte solution and $\mathcal{O}$ as the potential conjugate to the microscopic density. Clearly the partition function $\mathcal{Z}$ can be written exactly as Eq.~\ref{eq:Zeq} in terms of a functional integral,
\begin{equation}
    \mathcal{Z}[\mathcal{O}] = \int D[f] \exp{\left( -\beta \mathcal{H}[f; \mathcal{O}] \right)} = \exp{\left( -\beta F \right)}.
\end{equation}
Here we assume a bosonic scalar field restricted to a bilinear Hamiltonian,

\begin{widetext}
\begin{equation}
        \mathcal{H}[f;\mathcal{O}]
        = \frac{1}{2} \int_{\mathbb{R}^{2d}} d\xi d\xi'  f(\xi) V(\xi, \xi') f(\xi') - \int_{\mathbb{R}^d} d\xi f(\xi)\mathcal{O}(\xi),
\end{equation}
\end{widetext}
for some symmetric operator $V$. For example, we can choose $V = - \Delta + m^2$ where $\Delta = \nabla \cdot \nabla$ denotes the Laplacian operator on $\mathbb{R}^d$. Again taking the instance of $f$ describing the fluid density, $-\Delta$ then defines a kinetic term that captures the osmotic pressure while $\abs{m}$ defines a mass term that confines the fluid particles harmonically. Thus invoking the MF approximation on the space of product measures, we have
\begin{eqnarray}
    \beta F_{\rm MF} = \beta F + \inf_{\nu} \int D[f] \nu(f) \ln \left[ \frac{\nu(f)}{\rho_{\rm eq}(f)} \right],
\end{eqnarray}
with $\rho_{\rm eq}(f) \propto \exp{( -\beta \mathcal{H}[f] )}$. Now we want to point out that there are two complementary ways to define product measures $\nu$ here. Of course one way is to look at the measures such that for any integer $k \geq 1$,
\begin{eqnarray}
    \begin{split}
        \int D[f] \nu(f) \prod_{1\leq i \leq k} \delta (f(\xi_i)-f_i) \\
        &\hspace{-2.5cm} = \prod_{1 \leq i \leq k} \int D[f] \nu(f) \delta (f(\xi_i)-f_i),
    \end{split} 
\end{eqnarray}
where $\xi_{i} \in \mathbb{R}^d$ picks out the observation points and $\delta$ denotes the Dirac delta. In this case, we parametrize the free energy functional over the space of field averages, $\overline{f} = \langle f \rangle_{\nu}$, and deviations, $\sigma^2 = \langle f^2 \rangle_{\nu} - \langle f \rangle_{\nu}^2$. Let us take a trivial example of $V$ having a Green's function, $V^{-1}(\xi, \xi') = c \delta( \xi-\xi')$ for some $c >0$. We know that the exact theory is a MFT so we expect to retrieve the Gaussian measure $\rho_{\rm eq}$ as a result of optimizing over all possible $\overline{f}$ and $\sigma$. Adopting our previous notations, the MF solution can be derived by functional differentiations,
\begin{eqnarray}
        \frac{\delta \mathcal{F}_{\rm MF}[\overline{f},\sigma]}{\delta \overline{f}(\xi)} = \frac{\delta \mathcal{F}_{\rm MF}[\overline{f},\sigma]}{\delta \sigma(\xi)} = 0, 
        \label{eq:white_noise}
\end{eqnarray}
where $\mathcal{F}_{\rm MF} = \mathcal{H}_{\rm MF} - T  S_{\rm MF}$ is an infinite-dimensional generalization of the MF free energy function. Clearly the optimality condition above recovers Gaussian statistics where $\langle f(\xi) \rangle_{\nu_{\rm MF}} = \mathcal{O}(\xi)$ and $\langle f(\xi)f(\xi') \rangle_{\nu_{\rm MF}} = c k_{\rm B}T \delta(\xi-\xi')$. Although $\overline{f}$ and $\sigma$ appear independently in Eq.~\ref{eq:white_noise}, it is possible, as for the dipolar chain, that $\sigma = \sigma(\overline{f})$ if we consider fields subject to extra constraints. Then we get the MF equation,
\begin{eqnarray}
     \begin{split}
        \frac{\delta \mathcal{F}_{\rm MF}[\overline{f};V, \mathcal{O}]}{\delta \overline{f}(\xi)} = 0 \\ &\hspace{-1.25 cm} \implies \hat{\mathcal{G}}_{\rm MF}[\overline{f}](\xi) = \mathcal{Q}(\beta \hat{E}_{\rm MF}[\overline{f}](\xi)),
    \end{split}
    \label{eq: MF_field_eq}
\end{eqnarray}
where the operator $\hat{E}_{\rm MF}$ gives the effective MF local potential,

\begin{figure*}[t!]
\centering
\includegraphics[width=\textwidth]{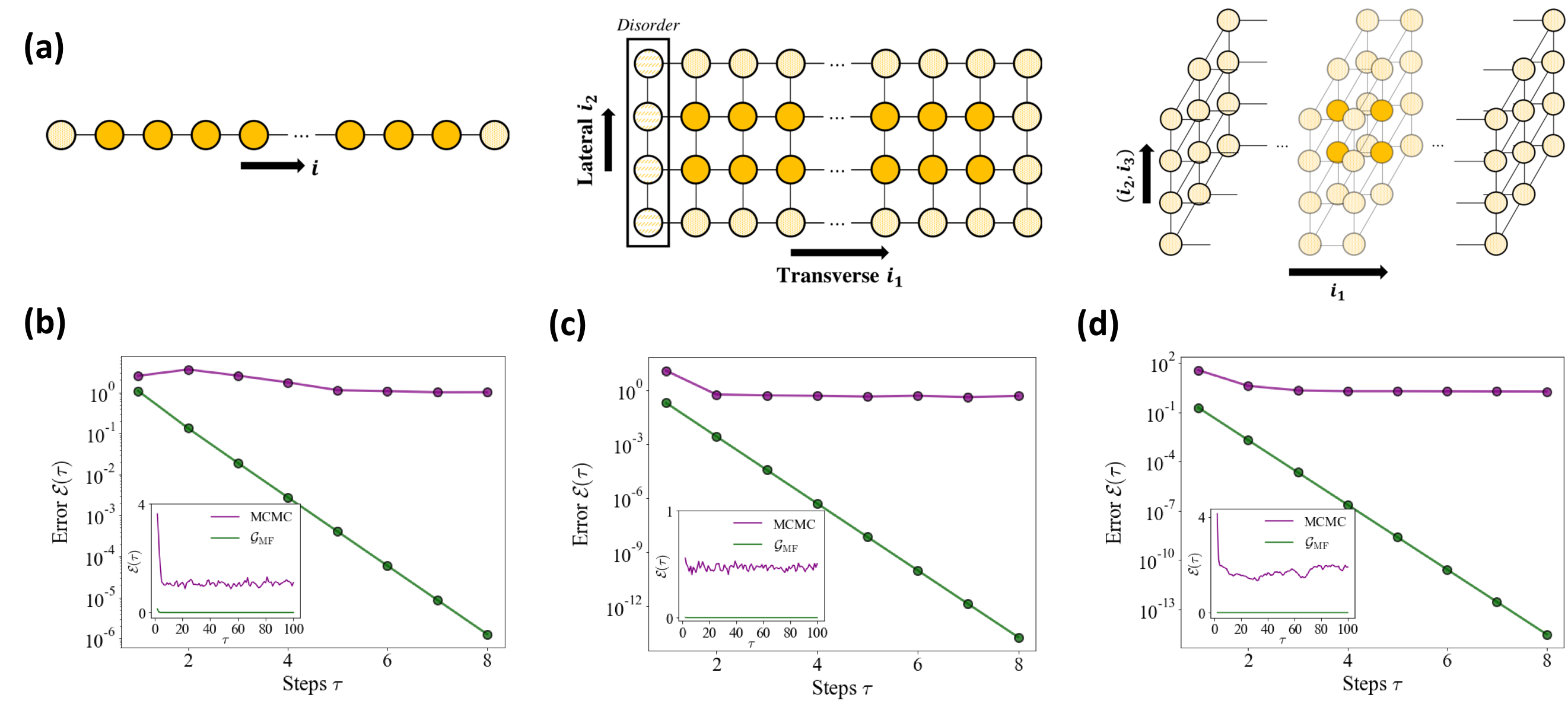}
\caption{ Numerical closure of MCMC and $\mathcal{G}_{\rm MF}$ iterations for spin systems of increasing dimensionality. $(a)$ Lattice geometry of dipolar chain, XY plaquette, and Heisenberg slab from left to right respectively. The boundary spins are distinguished from those in the interior by their shading. Error of convergence $\varepsilon(\tau)$ is plotted against the number of run steps $\tau$ for $(b)$ \textit{dipolar chain} $(c)$ \textit{XY plaquette} and $(d)$ \textit{Heisenberg slab}. Results from MC and $\mathcal{G}_{\rm MF}$ iterations are shown in purple and green respectively, where the asymptotic behaviors are displayed inset.
}
\label{fig:models}
\end{figure*}

\begin{eqnarray}
    \hat{E}_{\rm MF}[\overline{f}](\xi) = -\int_{\mathbb{R}^d} d\xi' V(\xi, \xi') \overline{f}(\xi') + \mathcal{O}(\xi), 
    \label{eq: MFint_1}
\end{eqnarray}
and $\mathcal{Q}$ is the activation function whose precise form is determined by the maximum entropy function associated with the random variable $f(0^d)$ centered at $\overline{f}(0^d)$. Note that Eq.~\ref{eq: MF_field_eq} can be recognized as the stationary limit of the Wilson-Cowan equation~\cite{Wilson-Cowan}. To avoid singularities, a cutoff of order $a$ in the configuration space or $2\pi/a$ in the frequency space is always implied in the integrals over $\mathbb{R}^d$, where we recall that $a$ sets the nearest neighbor distance on a lattice. The same prescription applies when we deal with fields supported on a compact subset to address finite system volume. This should not be too surprising since a functional integral is typically evaluated by a discretization of the field domain with a differential volume $a^d$. However, we see that sufficiently strong regularity of $V$ is required to make sense of Eq.~\ref{eq: MFint_1}, \textit{e.g.}, $V$ in the uncorrelated Gaussian model causes a somehow problematic interpretation.

We may turn to a different characterization of the product measures. In particular, we allow correlated fluctuations to occur over the configuration space, and instead look at induced measures from some unitary field transformation $U$ for which the bilinear term in $\mathcal{H}$ has a trivial kernel. In the example of $V^{-1}(\xi,\xi') = c \delta(\xi-\xi')$ above, $U$ is a Fourier transform, \textit{i.e.}, $f(\xi) \mapsto \Tilde{f}(\xi)$, where the integrals behave regularly at small wave numbers after a unitary rotation. The transformed fields $\Tilde{f} = \Re\Tilde{f} + i \Im\Tilde{f}$ are complex so the Hamiltonian becomes,
\begin{eqnarray}
    \begin{split}
        \mathcal{H} = \frac{1}{ c}\int_{\mathbb{H}^{d}} d\xi \left[\Re\Tilde{f}(\xi)^2 + \Im\Tilde{f}(\xi)^2 \right] \\
        &\hspace{-3 cm} - \int_{\mathbb{R}^d} d\xi \left[ \Re\Tilde{f}(\xi) - i\Im\Tilde{f}(\xi) \right] \Tilde{\mathcal{O}}(\xi),
    \end{split}
\end{eqnarray}
with an effective kernel $\Tilde{V}(\xi,\xi') = c^{-1} \delta(\xi-\xi')$. Let us assume $\mathcal{O}(\xi) \equiv 0$ to avoid further technicality. The equilibrium measure $D[\Re\Tilde{f}] D[\Im\Tilde{f}] \exp{(-\beta \mathcal{H}[\Re\Tilde{f}, \Im\Tilde{f}])}$ in this case may be regarded as a Gaussian measure on two real-valued fields over the upper half space $\mathbb{H}^d \subset \mathbb{R}^d$. We recover Gaussian statistics from the $U$-transformed MF optimization, where $\langle \Re\Tilde{f} \rangle_{\Tilde{\nu}_{\rm MF}} = \langle \Im\Tilde{f} \rangle_{\Tilde{\nu}_{\rm MF}} = 0$ and $2\langle \Re\Tilde{f}(\xi)\Re\Tilde{f}(\xi') \rangle_{\Tilde{\nu}_{\rm MF}} = 2\langle \Im\Tilde{f}(\xi)\Im\Tilde{f}(\xi') \rangle_{\Tilde{\nu}_{\rm MF}} = c k_{\rm B}T \delta(\xi-\xi')$. Note that it is easy to identify the equivalence of $\nu_{\rm MF}$ and $\Tilde{\nu}_{\rm MF}$ from the unitarity of Fourier transform. For general coupling $V$ and transformation $U$, a transformed MF equation of the same form as Eq.~\ref{eq: MF_field_eq} can be derived with proper analytic continuation.

We are interested in Gaussian measures because they satisfy the so-called reflection positivity condition~\cite{JaffeRP}. This special condition endows the algebra of classical fields a Hilbert space structure with well-defined vacuum state and field operators. It is hence hopeful to derive relevant results in quantum field theory with existing tools, although this is beyond the scope of the current work. To end the section, we want to comment that there is no straightforward extension of the convergence criteria for anisotropic models when the lattice $\Lambda$ belongs to an arbitrary crystal family in higher dimensions. In fact, a dipolar model in the absence of external field does not enthalpically favor a uniformly polarized configurations on a $d$-dimensional cubic lattice when $d \geq 2$.

\section{\label{sec:cost} Computational Cost and Stability}
We now present numerical data that demonstrate the utility of the iterative mapping $\mathcal{G}_{\rm MF}$. In our demonstration, we examine how the iterator performs across lattice systems of various dimensions, graphically represented in Fig.~\ref{fig:models}$(a)$. For comparison, we benchmark our results against those generated from empirical sampling using self-consistent Markov chain Monte Carlo (MCMC) ~\cite{SCMCMC, SCMCMC_Muller}. The MCMC scheme relies on progressive updates that modify the static spin environments until we approximately converge to the MF solution.

\subsection{1D dipolar chain}

We first revisit our minimal example of dipolar chain under free boundary conditions. We choose a uniform external field $E_{i} \equiv E_{\rm ext}$ and polarity variation $d_{i \in \partial \Lambda}$ $\neq d_{\rm bulk}$ only across the lattice boundary $\partial \Lambda = \{1,n\}$. We then analyze convergence of the MCMC and $\mathcal{G}_{\rm MF}$ schemes for system composed of $n=100$ dipoles. Specifically, the schemes are implemented by,
\begin{center}
\begin{algorithm}[H]
 \KwData{spin variables $ \underline{\vartheta} \in [-\pi,\pi)^{n}$}
 initialization of system replicas within a range of interpolated temperatures\;
 sampling a starting mean configuration $(\underline{\theta}_0, \underline{r}_0)$ at desired temperature through replica exchange\;
 $\tau = 0$\;
 \While{$\varepsilon(\tau) > \varepsilon_{\rm tol}$}{
      $\underline{\vartheta}_{\tau}, d_{\Lambda} \gets \underline{\theta}_{\tau}, \underline{r}_{\smash{\tau}}$\;
      \For{$i \gets 1$ \KwTo $n$}{
      $(\theta_{i,\tau}, r_{i,\tau}) \gets \left\langle d_i 
      \hat{\mu}_i \right\rangle$ via Metropolis updates with $(\vartheta_{j,\tau}, d_{j}) = (\theta_{j,\tau}, r_{j,\tau})$ fixed for $j \neq i$\;
      }
      $\underline{\theta}_{\tau+1}, \underline{r}_{\tau+1} \gets \underline{\theta}_{\tau}, \underline{r}_{\smash{\tau}}$\;
      $\tau \leftarrow \tau+1$\;
 }
 \caption{Empirical MCMC routine}
\end{algorithm}    
\end{center}
and 
\begin{center}
\begin{algorithm}[H]
\KwData{projected MF coordinates $\underline{X} \in \Delta_{\mathcal{D}}$}
 initialization of MF coordinates $\underline{X}_0$ \;
 $\tau = 0$\;
 \While{$\varepsilon(\tau) > \varepsilon_{\rm tol}$ }{
  $\underline{X}_{\tau+1} \leftarrow  \mathcal{G}_{\rm MF}(\underline{X}_{\tau}) $\;
  $\tau \leftarrow \tau+1$\;
 }
 \caption{Algebraic $\mathcal{G}_{\rm MF}$ routine}
\end{algorithm}    
\end{center}
where we define the single run step to be an MCMC sweep or a $\mathcal{G}_{\rm MF}$ recursion in the two schemes respectively. To quantify update progress, we define the error $\varepsilon$ over successive steps,
\begin{eqnarray}
    \hspace{-0.25cm} \varepsilon^2(\tau) = \sum_{i=1}^{n} d_i^2 \left[ (X_{i,\tau} - X_{i,\tau-1})^2+(Y_{i,\tau} - Y_{i,\tau-1})^2 \right],
\end{eqnarray}
where we set $Y_{i,\tau} = Y_{i,\tau-1} \equiv 0$ in the $\mathcal{G}_{\rm MF}$ scheme by default. Fig.~\ref{fig:models}$(b)$ shows the performance of the two schemes presented  above. We notice a rapid asymptotic decay of the convergence error in the algebraic scheme.  Within each run step, the parallelizability of the vectorized $\mathcal{G}_{\rm MF}$ operations can significantly save the actual runtime, although the computational complexity also depends on the system size $n$ for which both schemes share the same $O(n^2)$ scaling per step.

\begin{figure}
\centering
\includegraphics[width=.4\textwidth]{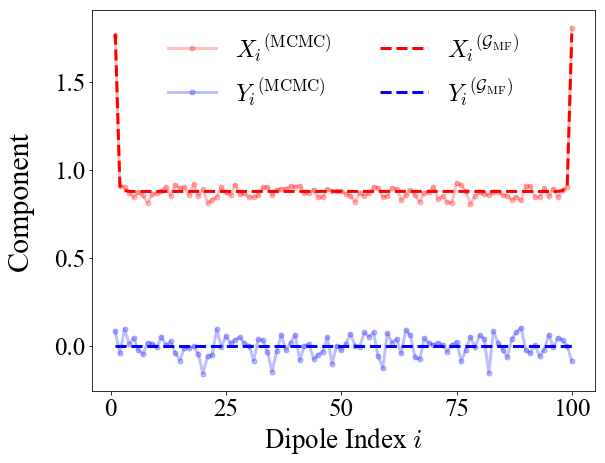}
\caption{Components of the MF polarization calculated from MCMC and $\mathcal{G}_{\rm MF}$ iterations with the model parameters $(\beta, E_{\rm ext}, d_{1}, d_{\rm bulk}, a) = (1, 0.2, 2, 1, 1)$. MF polarization profile is plotted along the 1D chain. Components extracted from MC and $\mathcal{G}_{\rm MF}$ are marked in solid and dashed lines respectively.
}
\label{fig:MCMC_Iteration}
\end{figure}

Figure~\ref{fig:MCMC_Iteration} displays the mean polarization profile across the dipolar chain up to a total of $500$ run steps, where listed model parameters are nondimensionalized by molecular units. The MCMC sampling is noisy due to constant trapping of the system near energy local minima. Although alternative MCMC strategies, such as cluster-based methods~\cite{clusterMC}, are well-suited for overcoming the sampling issue, they require additional computational resource to resolve systems that have extensive couplings $V_{ij}$ and break the lattice translational invariance. On the other hand, the $x$-component of MF polarization profile extracted from the $\mathcal{G}_{\rm MF}$ iterator precisely matches that recovered from prototypical message-passing inferences~\cite{MJ,NeuronMP} on the full mean polarization (discussed in Appendix C). Overall, we see that when applicable, $\mathcal{G}_{\rm MF}$ efficiently solves the MF model at a given accuracy.

\begin{figure}[b]
\centering
\includegraphics[width=.4\textwidth]{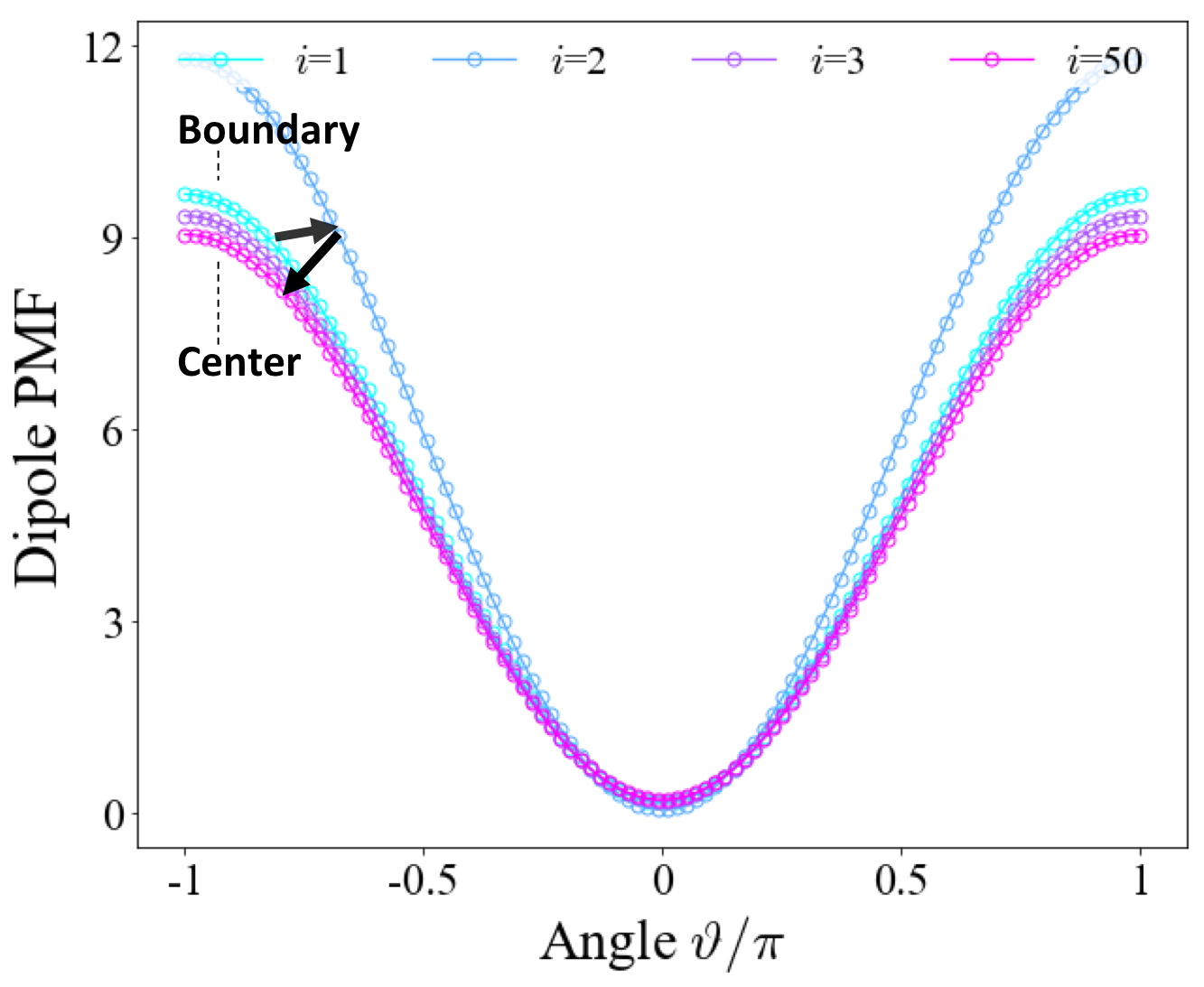}
\caption{Potential of mean force (PMF), $-k_{\rm B} T \ln \nu_i^{\star}$, are plotted for dipoles occupying different positions on the 1D chain. Marker color varies from light blue ($i=1$ at the boundary of $\Lambda$) to purple ($i=50$ at the center of $\Lambda$).}
\label{fig:MF_nus}
\end{figure}

From converged $\underline{X}_{\tau}$ under the $\mathcal{G}_{\rm MF}$-iterations, we recover the distributions $\nu_i^{\star}$ through Eq.~\ref{eq:nu_star}. The corresponding single-dipole MF statistics can be visualized in Fig.~\ref{fig:MF_nus}. We observe a nonmonotonic change in the thermodynamic force that drives the polarization response as we approach the lattice core from the boundary. Such persistent nonmonotonicity can be tuned as we alter characteristics of the polarity profile $d_{\Lambda}$. For example, if we modify the associated length scale in the polarity variation $i \mapsto d_i$ while fixing $d_1$ and $d_{\rm bulk}$ at $i=1$ and $i=n/2$ respectively, we effectively shift the population of dipoles that behave (statistically) like the "core" relative to those that behave like the "boundary".

\subsection{2D disordered interface}
To illustrate the application of the MF iterative approach to a different problem of physical relevance, we next consider a finite XY model (see Sec.~\ref{subsec:genearlized lattice models}) with distinct boundaries. We assume a square lattice $\Lambda = \{ (i_{1}, i_{2}): 1 \leq i_{k} \leq n_{k} \}$ which extends in two directions, a transverse direction describing the transition from an interface to the system interior as well as a lateral direction adopting additional spin heterogeneity. Instead of pushing the system with an external field, here we impose static lateral heterogeneity at one boundary layer, $i_1 = 1$, and maintain an open boundary condition at the other, $i_1=n_{1}$. Such heterogeneity could be realized, for example, if we randomize but freeze the orientation of the boundary spins, mimicking the quenched microscopic disorder at an interface.

\begin{figure}[H]
\centering
\includegraphics[width=.4\textwidth]{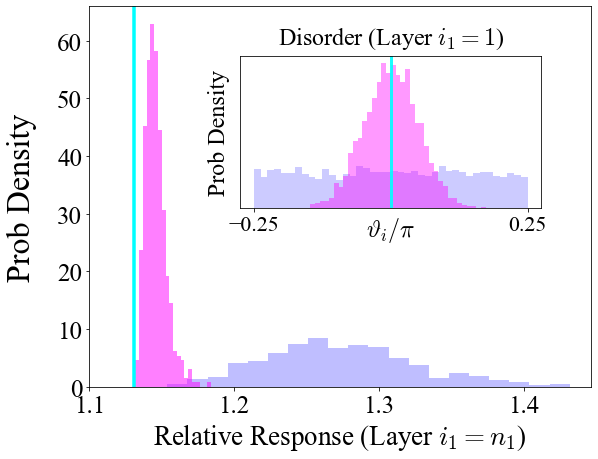}
\caption{Disorder-induced spin responses for 2D XY model with long-range interaction $V_{ij} \propto 1/r_{ij}$ and model parameters $(\beta, E_{\rm ext}, a, d_{i}) = (1, 0, 1, 1)$. At each transverse lattice position, spin response is measured as the variance of the MF distribution $ \nu_i^{\star}$ averaged over lateral direction. For a given disorder, response at the boundary layer $i_1 = n_1 = 20$ is reported in a histogram over a total of $500$ random realizations. The response is normalized by that at the interfacial layer $i_1 = 2$ and takes a value greater than $1$. The plot color distinguishes two different disorders (uniform and Gaussian), whose angular distributions are also shown inset. The cyan line marks the trivial case $\vartheta_i \equiv 0$.
}
\label{fig:MCMC_XY}
\end{figure}

For subsequent illustration, we first implement the MCMC and $\mathcal{G}_{\rm MF}$ schemes in the absence of any disorder, \textit{i.e.}, $\vartheta_i \equiv 0$ at the boundary layer $i_1 = 1$, and we exhibit the resulting rapid $\mathcal{G}_{\rm MF}$ convergence for a system of size $n=200$ in Fig.~\ref{fig:models}$(c)$. To explore the role of disorder, we randomly load static spins for which $-\pi/4 \leq \vartheta_i \leq \pi/4$ at the boundary layer $i_1 = 1$. In this case, the MF solution cannot be blindly reached via a $\mathcal{G}_{\rm MF}$ iteration, as our symmetry argument for $\theta_i^{\ast} = 0$ no longer holds due to the presence of static disorder (we lose our resolution over the $\underline{\theta}$ coordinate). However, up to a self-evident correction of the effective field which accounts for the orientational dependence on $\underline{\theta}$, the convergence properties of $\mathcal{G}_{\rm MF}$ are well-preserved. That is, we retain a rapid access to the constrained free energetic optimum under fixed $\underline{\theta}$. To this regard, we are able to port our iterator over, now as a free energy evaluation subroutine, to suitable global optimization routine, \textit{e.g.}, stimulated annealing and its variants~\cite{TSALLIS_1996, SA_1997, SA_2013}, for resolving $\underline{\theta}^{\ast}$ and thus the MF solution with high accuracy.

The influence of disorder may persistently extend from the quenched boundary into the system interior. The transverse spin response in Fig.~\ref{fig:MCMC_XY} to disorders at the interface is resolved using the hybrid method above (discussed in Appendix E). We notice that a change in the disorder at one boundary layer has a nontrivial impact on fluctuations of MF response at the other, where these calculations can benefit from the algorithmic acceleration from the $\mathcal{G}_{\rm MF}$ iteration.

\subsection{3D Heisenberg slab}

\begin{figure}[b]
\centering
\includegraphics[width=.425\textwidth]{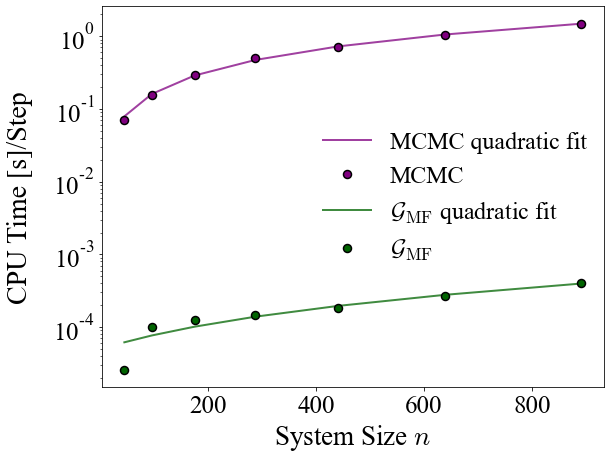}
\caption{System size scaling of MCMC and $\mathcal{G}_{\rm MF}$ iterations for 3D Heisenberg model with long-range interaction $V_{ij} \propto 1/r_{ij}$. Both schemes are run with the same set of parameters as from Fig.~\ref{fig:MCMC_Iteration}. Average single-core CPU time per run step is recorded for a series of system sizes $n$. The $O(n^2)$ scaling with respect to the system size is shown via solid curves (purple for MCMC and green for $\mathcal{G}_{\rm MF}$).
}
\label{fig:MCMC_Heisenberg}
\end{figure}

Finally, we evaluate the performance of the MF iterator on a Heisenberg spin system of comparable size. Specifically, we arrange the spins on a finite slab $\Lambda = \{ (i_{1}, i_{2}, i_{3}): 1 \leq i_{k} \leq n_{k} \}$ under uniform external field and consider, for simplicity, spin polarity variation only across the pair of parallel boundary surfaces $\partial \Lambda = \{(i_{1}, i_{2}, i_{3}): i_1 = 1~{\rm or}~i_1=n_{1} \} $. The convergence results are displayed in Fig.~\ref{fig:models}$(d)$, where the $\mathcal{G}_{\rm MF}$ iterator gives a rapid decay of error similar to those in Fig.~\ref{fig:models}$(b)$-$(c)$. Its efficiency is further demonstrated in Fig.~\ref{fig:MCMC_Heisenberg} marking the per-step runtime for increasingly large systems. The observed speedup of orders of magnitude suggests that the $\mathcal{G}_{\rm MF}$ scheme is capable of handling sizable heterogeneous systems.

\section{Discussion and Conclusion}
In this manuscript, we use a model of dipolar chain to motivate the mean field analysis of continuous spin models. In the infinite volume limit, the mean field approximation reduces to solving a single self-consistent equation that characterizes the bulk properties. When the system has finite volume and free boundaries, we use functional optimization to derive a condition that manifests the self-consistency of the resulting mean field equations, starting from a rudimentary thermodynamic variational principle. With consistent external fields, the mean field distribution and free energy profile can be rapidly constructed through a fixed-point iteration. Such mean field picture sheds light onto how individual spins orient under the average influence of each other, which minimally accounts for distinct bulk and interfacial solvent behaviors in the context of heterogeneous dipolar model, providing a statistical basis for studying interfacial dielectric response in driven electrochemical systems.

Our main results from Sections \ref{sec:iterator} and \ref{sec:genearlize} highlight the compatibility of the symmetry-based iterative approach, where the properties of a general class of mean field models can be retrieved from an optimization over the space of configurational probabilities. We restate the above infinite-dimensional nonconvex optimization problem as a finite-dimensional min-max problem that is locally convex. As a consequence, we are able to arrive at familiar mean field equations~\cite{MFdipole,MFsphere} available in the thermodynamic limit, with trivial modifications accounting for site heterogeneity. At the same time, we build up novel understanding about classical spin systems subject to external field. In fact, we have proven in Section \ref{sec:genearlize} that spherical spin chains with ferromagneticity and consistent external fields are all isomorphic, up to a renormalization of control parameters including the temperature, lattice spacing, and strength of external field. Sections \ref{sec:cost} demonstrates practicality of the iterative approach as guaranteed by the convergence theorems. 

Clearly, the optimization idea can be formalized beyond the mean field approximation, where the probability distributions of interest contain other structural or hierarchical features that allow a dimensional reduction of the search space. For example, we may follow equivalent arguments to realize the Bethe approximation~\cite{InformationPhysics,Koehler2019}, which becomes exact in the limit of strongly localized spin interactions. On the other hand, if we seek tight estimate of the actual equilibrium measure, $\rho_\mathrm{eq}$, we can feed measures parametrized under these simplifying approximations as input to the machinery of normalizing flows~\cite{NormalizingFlowSurvey,NormalizingFlowFreeEnergy}, developed in the field of generative deep learning and statistical inference, to target the true free energy minimum.

\section*{Data Availability}
The convergence data in Section ~\ref{sec:cost} that support the findings of this work are available from the authors upon reasonable request.

\section*{Appendix}
\appendix
\section{Variational Optimization}
\subsection*{Graphical model representation}
Consider the system of interacting dipoles in the context of structured probabilistic models by formally associating a collection of $n$ point dipoles with an undirected graph, $\Lambda = (\mathcal{V}, \mathcal{E})$, whose vertex set $\mathcal{V}$ and edge set $\mathcal{E}$ encode the system interaction topology. Due to the nonlocal nature of electrostatics, we focus on graphs that are fully connected, \textit{i.e.}, $(i,j) \in \mathcal{E}$ for all $i \neq j$. Our model also assumes that the graph possesses a 1D lattice structure for which $\hat{x}: i \mapsto i+1$ acts as a primitive translation, inducing a distance $r_{ij} = a\abs{i-j}$ on the edge set for lattice spacing $a$. 

We denote the dipolar orientation by a vector $\underline{\vartheta} \in [-\pi,\pi)^n$ that specifies the angle each dipole makes with some reference direction, \textit{e.g.}, $\hat{\boldsymbol{x}} = (1,0)$. The energetic cost of a reorientation $\underline{\vartheta} \mapsto \underline{\vartheta} + \Delta \underline{\vartheta}$ is determined by the Hamiltonian,
\begin{eqnarray}
    \mathcal{H}( \underline{\vartheta} | d_{\Lambda}) &=& \sum_{(i,j) \in \mathcal{E}} V_{ij}(\vartheta_i, \vartheta_j) + \sum_{i \in \mathcal{V}} h_i(\vartheta_i), 
\end{eqnarray}
where the two-body potential $V_{ij}$ accounts for the pairwise interactions and the one-body potential $h_{i}$ accounts for the external electric field. For simplicity, we look at planar angular fluctuations sufficient for our analysis of dielectric response. 

\subsection*{Free energy optimization}
The free energy of the system is,\begin{eqnarray}
    F = \min_{\rho \in \mathcal{P}(\underline{\vartheta})} \mathcal{F}[\rho] = \min_{\rho \in \mathcal{P}(\underline{\vartheta})} \left[\langle\mathcal{H}\rangle_{\rho} - TS_{\rho} \right],
    \label{eq:Fmin}
\end{eqnarray}
where $\mathcal{P}(\underline{\vartheta})$ denotes the space of probability measures over configurations $[-\pi, \pi)^n$. Eq.~\ref{eq:Fmin} recapitulates the result from classical thermodynamics that the Boltzmann distribution optimally regulates energy fluctuations of canonical ensemble by achieving the global minimum of the convex functional $\mathcal{F}$. 

In general, the solution to MFT can be formulated as a constrained optimization problem,
\begin{eqnarray}
    \displaystyle \beta F_{\rm MF} = \beta F + \min_{\nu \in \mathcal{M}(\underline{\vartheta}) } D_{\rm KL}(\nu || \rho_{\rm eq}) \geq \beta F,
    \label{eq:MF_var}
\end{eqnarray}
where $F_\mathrm{MF}$ and $F$ designate the free energies of the mean field and full systems respectively, and $\mathcal{M}(\underline{\vartheta})$ denotes the search space of fully factorizable measures $\nu(\underline{\vartheta}) = \prod_{i=1}^n \nu_i(\vartheta_i)$ over the many-body configuration space.
The entropic penalty associated with the mean field construction is captured by the Kullback–Leibler (KL) divergence,
\begin{eqnarray}
     & \displaystyle D_{\rm KL}(\nu || \rho_{\rm eq}) = \int d\underline{\vartheta} \nu(\underline{\vartheta}) \ln \left[ \frac{\nu(\underline{\vartheta})}{\rho_{\rm eq}(\underline{\vartheta})} \right], \\
    &\displaystyle = \int d\underline{\vartheta} \nu(\underline{\vartheta}) \beta \mathcal{H}(\underline{\vartheta}) - k_{\rm B}^{-1} \sum_{i=1}^{n} S_{\nu_i}  + \text{constant},
    \label{eq:KL}
\end{eqnarray}
where $S_{\nu_i}$ denotes the entropy that encodes one-body fluctuations in the marginal distribution $\nu_i$. With $\mathcal{F}_{\rm MF}$ defined in the main text, the optimization problem posed from Eq.~\ref{eq:MF_var} can be simply re-expressed as an equivalent problem over a compact set $\Omega_0$ (which is the product of closed disks of radius $r_i$), 
\begin{eqnarray}
    F_{\rm MF} &=& \min_{\nu \in \mathcal{M}(\underline{\vartheta})} \mathcal{F}[\nu], \\
    &\equiv& \min_{\lbrace \Vec{u}_i \rbrace_{i=1}^{n} \in \Omega_0} \inf_{\nu \in \mathcal{M}(\underline{\vartheta}): \langle \hat{\mu}_i \rangle_{\nu} = \vec{u}_i }  \mathcal{F}[\nu],\\
    &=& \min_{\lbrace \Vec{u}_i \rbrace_{i=1}^{n} \in \Omega_0} \mathcal{F}_{\rm MF}(\underline{\theta},\underline{r}| d_{\Lambda}),
    \label{eq:F_MF_general}
\end{eqnarray}
so the minimizer $\nu_{\rm MF}$ attaining $\mathcal{F}[\nu_{\rm MF}] = F_{\rm MF}$ corresponds to an interior critical point $(\Vec{u}_1^{\ast}, \cdots, \Vec{u}_n^{\ast})$ of the MF free energy function, \textit{i.e.}, $\nabla_{\Vec{u}_i} \mathcal{F}_{\rm MF}(\underline{\theta}^{\ast}, \underline{r}^{\ast}) = 0$ (otherwise $(\Vec{u}_1^{\ast}, \cdots, \Vec{u}_n^{\ast}) \in \partial \Omega_0$ implies $r_i^{\ast} = d_i$ and thus $S_{\nu_{\text{MF}, i}} = -\infty$ for some $i$). 

\section{Proofs of Lemmas and Theorems}
Let $\Theta^{+} = [-\pi/2, \pi/2]$ denote the half angular window  and $\mathcal{D} = [0, d_1) \times \cdots \times [0, d_n)$ the space of mean polarizations.

\noindent \textit{{Proof}} of \textbf{Lemma 1.1.} For any $\mathcal{V}' \subset \mathcal{V}$, let $\mathcal{R}_{\mathcal{V}'}: [-\pi, \pi)^n \rightarrow [-\pi, \pi)^n$ be the partial reflection,
\begin{eqnarray}
    \left[ \mathcal{R}_{\mathcal{V}'}(\underline{\theta}) \right]_{i} = \begin{cases} [[ \pi - \theta_i ]] &\mbox{if } i \in \mathcal{V}', \\
    \theta_i &\mbox{otherwise}, \\
    \end{cases}
\end{eqnarray}
where the double bracket formally keeps $[[ s ]] \in [-\pi, \pi)$ upon shift of $2\pi \mathbb{Z}$. The isometry above trivially induces an entropy-preserving map $\nu^{\star}(\underline{\vartheta}|\underline{\theta}, \underline{r}) \mapsto \nu^{\star}(\underline{\vartheta}|\mathcal{R}_{\mathcal{V}'}(\underline{\theta}), \underline{r})$, as entropy is invariant under an orthogonal coordinate transformation. Now let $\mathcal{V}' = \{ i \in \mathcal{V}: \theta_i \in \Theta^{+} \} \subset \mathcal{V}$ and consider the partially reflected coordinates $(\underline{\theta}^{+}, \underline{r}) =  (\mathcal{R}_{\mathcal{V}- \mathcal{V}'}( \underline{\theta}), \underline{r})$.

If $\mathcal{V}'$ is empty, $V_{ij}(\theta^{+}_i, \theta^{+}_j) = V_{ij}(\theta_i, \theta_j)$ for all pairs $ (i,j) \in \mathcal{E}$ by symmetry of the dipolar interaction and $h_{i}(\theta^{+}_i) \leq h_{i}(\theta_i)$ for $i \in \mathcal{V}$. Otherwise, $V_{ij}(\theta^{+}_i, \theta^{+}_j) = V_{ij}(\theta_i, \theta^{+}_j) \leq V_{ij}(\theta_i, \theta_j)$ for pairs $(i,j) \in  ( \mathcal{V}', \mathcal{V} - \mathcal{V}')$ and $h_{i}(\theta^{+}_i) \leq h_{i}(\theta_i)$ for $i \in \mathcal{V} - \mathcal{V}'$. In either case, we see $\mathcal{H}_{\rm MF}(\underline{\theta}^{+}, \underline{r}) \leq \mathcal{H}_{\rm MF}(\underline{\theta}, \underline{r}) $ and thus $(\underline{\theta}^{+}, \underline{r})$ beats $(\underline{\theta}, \underline{r})$ as the maximizing candidate. By continuity of $\mathcal{F}_{\rm MF}$, the existence of a global maximizer $(\underline{\theta}^{\ast}, \underline{r}^{\ast})$ is ensured since $\Theta^{+} \times \mathcal{D}$ is compact. $\square$

\bigskip

\noindent \textit{{Proof}} of \textbf{Theorem 2.1.} We show that our statement follows from Banach-Caccioppoli contraction mapping theorem~\cite{FixedPointTheory}, arguably the most elementary yet versatile principle in fixed-point theory, by establishing a uniform upper bound $\mathcal{B}_{\nabla}$ on the derivatives $\nabla_{j} \mathcal{G}_{{\rm MF}, i}$. For $\underline{X} \in \mathcal{D}$ and $i \in \mathcal{V}$, let $\mathcal{V}^{i+}(\underline{X}) \subset \mathcal{V}-\{i\}$ denote the index set for which the mean polarity is positive and $\underline{R}_{i} \in \mathbb{R}^n$ with $(R_{i})_{j} = -\doubleunderline{T}^{ij}_{11}$ absorb the distance dependence of dipolar coupling for dipole $i$. 

For convenience, we define $\underline{X}^{i+}, \underline{R}_i^{i+} \in \mathbb{R}^{|\mathcal{V}^{i+}(\underline{X})|}$ to drop out ``boring" entries that involve no mean polarizations and therefore contribute vanishing mean fields. We will set the convention $\underline{X}^{i+} \equiv 0^{n-1}$ if $|\mathcal{V}^{i+}(\underline{X})| = 0$ and use the Kantorovich inequality to bound the inner product $\underline{R}_i^{\top} \underline{X}$,
\begin{eqnarray}
    \beta d_i E_{{\rm MF},i}(\underline{X}; \mathbf{E}^{\rm ext}_i) &\geq&  \beta d_i \left[  E_i  + B {\|\underline{R}_i^{i+}\|}_2 {\|\underline{X}^{i+}\|}_2 \right], \\
    &=& B_i(\underline{X};E_i), 
    \label{eq:l2_bound_specific}
\end{eqnarray}
where $B(\underline{X};i) = \lim_{\epsilon \rightarrow 0^{+}} 2\sqrt{b_1 b_0}/(b_1 + b_0 + \epsilon)$ is coordinate dependent with $b_1(\underline{X};i) = \min_{j \in \mathcal{V}} X^{i+}_{j}/n^{3}$ and $b_0(\underline{X}) = \norm{\underline{X}}_{\infty}$. Since the monotonicity of $\nabla_t q(t)$ implies,
\begin{eqnarray}
    \nabla_{j} \mathcal{G}_{{\rm MF},i}(\underline{X}) &\leq& \beta d_i^2 (R_i)_{j} \nabla_t q \left( B_i \right), \\
    &=& \beta d_i^2 (R_i)_{j} \left[ 1 - \frac{q(B_i)}{B_i} - q(B_i)^2 \right],
\end{eqnarray}
it suffices to bound the expression from the second equality more precisely in order to extract a uniform gradient bound $\mathcal{B}_\nabla$. Here we use the fact~\cite{BesselProperties}, 
\begin{eqnarray}
    \begin{split}
        \hspace{-0.5 cm} q(t) = \frac{I_{1}(t)}{I_{0}(t)} \geq \frac{t}{1+\sqrt{1+t^2}} & \\
         & \hspace{-2cm} \implies 1 - \frac{q(t)}{t} - q(t)^2 \leq  \frac{1}{1+\sqrt{1+t^2}},
    \end{split}
\end{eqnarray}
where $I_1(t) = \nabla_t I_{0}(t)$ gives the first order modified Bessel function. This allows the inequality,
\begin{eqnarray}
    \begin{split}
        \sup_{\underline{X} \in \mathcal{D}} \sup_{(i,j) \in \mathcal{E}} \abs{\nabla_{j} \mathcal{G}_{{\rm MF},i}(\underline{X})} \\
        &\hspace{-1.5cm} \leq \max_{i \in \mathcal{V}} \frac{2\beta d_i^2}{1+\sqrt{1 + (\beta d_i E_i)^2 }} = \mathcal{B}_{\nabla},
    \end{split}
\end{eqnarray}
which certainly implies $\mathcal{B}_{\nabla} < 1/n$ given that $E_i \geq E_i^{\ast} \geq 2n\norm{d_{\Lambda}}_{\infty}$, $\forall i \in \mathcal{V}$. For $E \geq E^{\ast}$, iterator $\mathcal{G}_{\rm MF}$ gives a contraction since $\forall \underline{X}',\underline{X}'' \in \mathcal{D}$,
\begin{eqnarray}
   \begin{split}
       \norm{ \mathcal{G}_{\rm MF}(\underline{X}') - \mathcal{G}_{\rm MF}(\underline{X}'')}_p \\
       &\hspace{-1.5cm} \leq \sup_{\lambda \in [0,1]} \norm{d\mathcal{G}_{\rm MF}\vert_{\underline{\rchi}_{\lambda}}}_p \norm{\underline{X}'-\underline{X}''}_p,
   \end{split}
\end{eqnarray}
where the multivariate mean value inequality with respect to the $\ell^{\hspace{0.02cm} p}$-norm for $1 \leq p \leq \infty$ derives from application of the fundamental theorem of calculus along an interpolating path $\underline{\rchi}_{\lambda} = \lambda \underline{X}' + (1-\lambda)\underline{X}''$ with $\lambda \in [0,1]$, \textit{i.e.},
\begin{eqnarray}
    \mathcal{G}_{\rm MF}(\underline{X}') - \mathcal{G}_{\rm MF}(\underline{X}'') = \int_{[0,1]} d\lambda d\mathcal{G}_{\rm MF}\vert_{\underline{\rchi}_{\lambda}} \cdot \nabla_{\lambda} \underline{\rchi}_{\lambda},
\end{eqnarray}
and the operator norm $\norm{d\mathcal{G}_{\rm MF}}_p \leq \norm{d\mathcal{G}_{\rm MF}}_1^{1/p} \norm{d\mathcal{G}_{\rm MF}}_{\infty}^{1-1/p}  \leq n \mathcal{B}_{\nabla} \leq \mathcal{B}$ is bounded below unity according to the Riesz-Thorin theorem~\cite{RieszThorin}. So by the continuity of norm as well as triangle inequality, we indeed observe a linear $\ell^{\hspace{0.02cm} p}$-convergence towards the fixed-point $\underline{X}^{\ast}$,
\begin{eqnarray}
    \hspace{-0.85 cm} \norm{\mathcal{G}_{\rm MF}^{(\tau)}(\underline{X}_0) - \underline{X}^{\ast} } &\leq& \lim_{s \rightarrow \infty} \sum_{k=\tau}^{\tau+s} \norm{\mathcal{G}_{\rm MF}^{(k)}(\underline{X}_0) - \mathcal{G}_{\rm MF}^{(k+1)}(\underline{X}_0) }, \\
    &\leq& \sum_{k=\tau}^{\infty} \mathcal{B}^{k} \norm{\mathcal{G}_{\rm MF}(\underline{X}_0)- \underline{X}_0},
\end{eqnarray}
where the last inequality yields a convergence factor $\mathcal{B}_{\tau}$ as a geometric series. Here we will omit the proof of existence and uniqueness of $\underline{X}^{\ast}$, which directly follows from the contractive property of the iterator.  We want to make a quick remark that Eq.~\ref{eq:l2_bound_specific}, although not explicit in the gradient bound for the case $a=1$, is sharp in the sense that it leads to alternative forms of convergence factor when we vary specific model parameters and restrict the domain of convergence, \textit{e.g.}, $2n \max_{i} d_i / E^{\ast}_i \gg 1$ while $d_{i} \gg 1$ and $a \geq 1$. $\square$

\bigskip

\noindent \textit{{Proof}} of \textbf{Lemma 2.2.}  Suppose that two fixed points, $\underline{X}_c$ and $\underline{X}_c'$, of $\mathcal{G}_{\rm MF}$ exist. By convexity of the hypercube $\bigtimes_{i \in \mathcal{V}} [0, d_i]$, an interpolating path of the form,
\begin{eqnarray}
    \underline{\rchi}_{\lambda} = \lambda \underline{X}_c + (1-\lambda) \underline{X}_c';~\lambda \in [0,1],
\end{eqnarray}
remains inside the hypercube. Moreover, the interpolation extends to boundary of the half-open rectangle $[0, \infty)^n$ on some $[\lambda_{-}, \lambda_{+}] \supset [0,1]$. Let us assume without loss of generality that $ {(\underline{X}_c')}_i < {(\underline{X}_c)}_i$ for some $i \in \mathcal{V}$ and define
\begin{eqnarray}
    I(\lambda, i) = \mathcal{G}_{{\rm MF}, i}(\underline{\rchi}_{\lambda}) - {(\underline{\rchi}_{\lambda})}_{i},
\end{eqnarray}
where $I(0,i) \equiv I(1,i) = 0$ from our construction. The function $I(\lambda, i)$ analytically continues onto the half-open rectangle so we have $(\underline{\rchi}_{\lambda_{-}})_{i} = 0$ and $I(\lambda_{-}, i) > 0$ (otherwise consider the vertex with the earliest hitting time $ \argmin_{j \in \mathcal{V}} (\underline{\rchi}_{\lambda_{-}})_{j}$). From the mean value theorem, we can find $\lambda_0 \in (0,1)$ and $\lambda_< \in (\lambda_{-},0)$ such that $\nabla_{\lambda} I(\lambda_0, i) = 0$ and $\nabla_{\lambda} I(\lambda_<, i) < 0$. However, note that
\begin{eqnarray}
    \nabla_{\lambda} I(\lambda, i) &=& \left[ \nabla \mathcal{G}_{{\rm MF},i}(\underline{\rchi}_{\lambda}) \right]^{\top}\delta \underline{X}_c - (\delta \underline{X}_c)_i, \label{eq:critical_pt_grad} \\
    \nabla_{\lambda} \nabla_{\lambda} I(\lambda, i) &=& \delta \underline{X}_c^{\top}
    \nabla \nabla \mathcal{G}_{{\rm MF},i}(\underline{\rchi}_{\lambda}) \delta \underline{X}_c,
    \label{eq:critical_pt}    
\end{eqnarray}
for which $\delta \underline{X}_c = \underline{X}_c - \underline{X}_c'$ and $\nabla \nabla \mathcal{G}_{{\rm MF},i}(\underline{X}) \in \mathbb{R}^{n \times n}$ with entries $ \nabla_j \nabla_k \mathcal{G}_{{\rm MF},i} = \beta^2 d_i^3 \nabla_t (R_{i})_{j} (R_i)_{k} \nabla_t q(\beta d_i E_{{\rm MF},i})$ is negative semidefinite since $\nabla_t \nabla_t q(t) < 0$ for $t > 0$. This leads to a contradiction since Eq.~\ref{eq:critical_pt} implies $\nabla_{\lambda} I(0, i) \geq \nabla_{\lambda} I(\lambda_0,i) = 0$ while $\nabla_{\lambda} I(0, i) \leq \nabla_{\lambda} I(\lambda_<, i) < 0$. $\square$

\bigskip

\bigskip

\noindent \textit{{Proof}} of \textbf{Theorem 2.3.} Let $\Delta_{\mathcal{D}} = \{\underline{X} : \underline{X} \geq \underline{X}^{\ast} \}$ with $``\geq"$ defined entry-wise. Stability of $\Delta_{\mathcal{D}}$ follows from the simple observation $\nabla_t q(t) > 0$ for $t \geq 0$, since
\begin{eqnarray}
    \underline{X}_{\tau} \geq \underline{X}^{\ast} \implies E_{{\rm MF},i}(\underline{X}_{\tau}) \geq E_{{\rm MF},i}(\underline{X}^{\ast}),
\end{eqnarray}
implies $(\underline{X}_{\tau+1})_i = d_i q(\beta d_i E_{{\rm MF},i}) =  \mathcal{G}_{{\rm MF},i}(\underline{X}_{\tau}) \geq \mathcal{G}_{{\rm MF},i}(\underline{X}^{\ast}) = X^{\ast}_i$ for $i \in \mathcal{V}$.

Next we show that free energy Hessian $\nabla_{\underline{X}} \nabla_{\underline{X}} \Phi_{\rm MF}(\underline{X}, 0^n)$ with entries $\nabla_{j} \nabla_{k} \Phi_{\rm MF} = \beta^2 \langle \delta E_{{\rm tot}, j} \delta E_{{\rm tot}, k} \rangle_{\underline{X}}$ stays negative semi-definite over the convex region $\Delta_{\mathcal{D}}$. Here $\langle ~\cdot~ \rangle_{\underline{X}}$ denotes the expectation under a fixed collective mean polarization $\underline{X}$, and $\delta E_{\rm tot} = E_{\rm tot} - \langle E_{\rm tot} \rangle_{\underline{X}}$ denotes fluctuations of the effective net field on individual dipoles. Note that $\forall \underline{X} \in \Delta_{\mathcal{D}}$,
\begin{eqnarray}
    \begin{split}
        \nabla_{\underline{X}} \nabla_{\underline{X}} \Phi_{\rm MF}(\underline{X}) - \nabla_{\underline{X}} \nabla_{\underline{X}} \Phi_{\rm MF}(\underline{X}^{\ast}) \\
        &\hspace{-5cm} \displaystyle = \underset{i \in \mathcal{V}}{{\rm Diag}} \left[ \frac{1}{d_{i}^2 \nabla_t q\big(q^{-1}\left(\frac{X_i^{\ast}}{d_i}\right)\big)} - \frac{1}{d_{i}^2 \nabla_t q\big(q^{-1}\left(\frac{X_i}{d_i}\right)\big)} \right],
    \end{split}
    \label{eq:Hessian_MF}
\end{eqnarray}
where the RHS is negative semidefinite because its diagonals are non-positive (recall $\nabla_t q^{-1}(t) > 0$ and $\nabla_t \nabla_t q(t) < 0$ for $0< t \leq 1$). By Eq.~\ref{eq:Hessian_MF}, $\nabla_{\underline{X}} \nabla_{\underline{X}} \Phi_{\rm MF}|_{\Delta_{\mathcal{D}}}$ is negative semidefinite when $\underline{X}^{\ast} \in \mathcal{D} - \partial \mathcal{D}$, and it suffices to examine the statement when $\underline{X}^{\ast} \in \partial \mathcal{D}$. We argue heuristically that the regime $X^{\ast}_i = 0$ for any site $i \in \mathcal{V}$ is physically irrelevant because the spin-spin couplings and external field tend to align the individual dipoles along some direction at finite temperatures. Otherwise, a direct computation shows $\underline{X}^{\ast} \in \mathcal{D} - \partial \mathcal{D} \implies \underline{X}^{\ast} = 0^n$ only in the limit $E \rightarrow 0^n$.

Using the curvature condition above, we have
\begin{eqnarray}
     \Phi_{\rm MF}(\underline{X}^{\ast}) &-& \Phi_{\rm MF}( \mathcal{G}^{(\tau)}_{\rm MF}(\underline{X}_{0})) \nonumber \\ 
     &=& -\int_{[0,1]} d\lambda \nabla_{\lambda} \Phi_{\rm MF}(\underline{\rchi}_{\lambda}^{\ast}), \label{eq:error} \\
     &\leq& \nabla_{\underline{X}} \Phi_{\rm MF}(\underline{X}_{\tau})^{\top} \delta \underline{X}_{\tau} \int_{[0,1]} d\lambda \lambda, \label{eq:curvature_condition}
\end{eqnarray}
if we consider an interpolating path $\underline{\rchi}_{\lambda}^{\ast} = \lambda \underline{X}_{\tau} + (1-\lambda) \underline{X}^{\ast}$ and apply the fundamental theorem of calculus to arrive at the integral in the second equality above. Here $\delta \underline{X}_{\tau} = - \nabla_{\lambda} \underline{\rchi}_{\lambda}^{\ast} = \underline{X}^{\ast} - \underline{X}_{\tau}$ and Eq.~\ref{eq:curvature_condition} follows, for $\lambda_2 \leq \lambda_1$, from
\begin{eqnarray}
    \hspace{-0.5cm} \begin{cases}
    \nabla_{\lambda} \Phi_{\rm MF}(\underline{\rchi}_{\lambda_1}^{\ast}) \leq \nabla_{\lambda} \Phi_{\rm MF}(\underline{\rchi}_{\lambda_2}^{\ast})  \leq  \nabla_{\lambda} \Phi_{\rm MF}(\underline{\rchi}_{0}^{\ast}), \\
    \\
    \nabla_{\lambda} \nabla_{\lambda} \Phi_{\rm MF}(\underline{\rchi}_{\lambda_1}^{\ast}) \leq \nabla_{\lambda} \nabla_{\lambda} \Phi_{\rm MF}(\underline{\rchi}_{\lambda_2}^{\ast}) \leq 0,
    \end{cases}
\end{eqnarray}
due to local concavity of the free energy surface $(\underline{X}, \Phi_{\rm MF}(\underline{X}))$. Notice that the vector $\delta\underline{X}_{\tau}$ contains exclusively non-positive entries all bounded above in magnitude by $\norm{d_{\Lambda}}_{\infty}$. For $\underline{X}_0 \in \Delta_{\mathcal{D},0} = \{ \underline{X} \in \bigtimes_{i \in \mathcal{V}} (0,d_i]: \underline{X} \geq \mathcal{G}_{\rm MF}(\underline{X}) \} \cap \Delta_{\mathcal{D}}$, the free energy gradient also reserves non-positive entries when evaluated at $\underline{X}_{\tau}$, \textit{i.e.}, 
\begin{eqnarray}
     \hspace{-2cm} && \displaystyle \norm{\nabla_{\underline{X}} \Phi_{\rm MF}(\underline{X}_{\tau})}_{1} = -\sum_{i \in \mathcal{V}} \nabla_i \Phi_{\rm MF}(\underline{X}_{\tau}),
     \label{eq:L1_gradient}
     \\
     &=& - \sum_{i \in \mathcal{V}} \beta E_{{\rm MF},i}(\underline{X}_{\tau}) - d_{-i} q^{-1} ( d_{-i}(\underline{X}_{\tau})_i ), \\
     &=& \beta \sum_{i \in \mathcal{V}} E_{{\rm MF},i}(\underline{X}_{\tau-1}) - E_{{\rm MF},i}(\underline{X}_{\tau}), \\
     &=& \beta \sum_{i \in \mathcal{V}} \Delta E_{{\rm MF},i}^{(\tau)} ,
     \label{eq:telescope}
\end{eqnarray}
where Eq.~\ref{eq:L1_gradient} reflects the simple observation $\underline{X}_{0} \geq \underline{X}_{1} \implies \underline{X}_{\tau} \geq \underline{X}_{\tau+1} \iff \nabla_i \Phi_{\rm MF} (\underline{X}_{\tau}) \leq 0$ for $i \in \mathcal{V}$. The subregion $\Delta_{\mathcal{D},0}$ is nonempty since the iterator has strictly bounded components, \textit{i.e.}, $ d_i > d_i q(\beta d_z E_{{\rm MF},i}(d_{\Lambda})) \geq \mathcal{G}_{{\rm MF},i}$, and the cascade of inequalities $\underline{X}_{\tau} \geq \underline{X}_{\tau+1}$ due to such a choice of region of initial estimates further implies,
\begin{eqnarray}
    \Phi_{\rm MF}(\underline{X}_{\tau+1}) \geq \Phi_{\rm MF}(\underline{X}_{\tau}),
    \label{eq:F_monotone}
\end{eqnarray}
if we consider interpolations $\lambda \underline{X}_{\tau} + (1-\lambda) \underline{X}_{\tau+1}$ and apply the fundamental theorem of calculus to the line integral,
\begin{eqnarray}
    \begin{split}
         \hspace{-0.5cm} \Phi_{\rm MF}(\underline{X}_{\tau}) - \Phi_{\rm MF}(\underline{X}_{\tau+1}) \\
         &\hspace{-1.5cm} = \int_{[0,1]}  d\lambda  \nabla_{\lambda} \Phi_{\rm MF}(\lambda) \leq \nabla_{\lambda} \Phi_{\rm MF}(0),
    \end{split}
\end{eqnarray}
where the RHS is nothing but $\nabla \Phi_{\rm MF}(\underline{X}_{\tau+1})^{\top} [ \underline{X}_{\tau} - \underline{X}_{\tau+1} ] \leq 0$. Combining Eqs.~\ref{eq:curvature_condition}, \ref{eq:telescope}, and \ref{eq:F_monotone}, we can bound the iterator error in terms of a telescoping series,
\begin{eqnarray}
    \hspace{-1 cm} && \abs{\Phi_{\rm MF}( \mathcal{G}^{(\tau)}_{\rm MF}(\underline{X}_{0})) - \Phi_{\rm MF}(\underline{X}^{\ast})} \nonumber \\
    &\leq& \frac{1}{\tau} \sum_{m=1}^{\tau}  \Phi_{\rm MF}(\underline{X}^{\ast}) - \Phi_{\rm MF}(\underline{X}_{m}), \\
    &\leq& \frac{\beta \norm{d_{\Lambda}}_{\infty}}{2\tau} \sum_{m=1}^{\tau} \sum_{i \in \mathcal{V}} \Delta E_{{\rm MF},i}^{(m)}, \\
    &=& \frac{\beta \norm{d_{\Lambda}}_{\infty}}{2\tau} \sum_{i=1}^{n} E_{{\rm MF},i}(\underline{X}_0) - E_{{\rm MF},i}(\underline{X}_{\tau}), \\
    &\leq& \frac{\beta \norm{d_{\Lambda}}_{\infty}}{\tau}
    \sum_{k=1}^{[n/2]} n k^{-3} \norm{d_{\Lambda}}_{\infty},
\end{eqnarray}
where $[~\cdot~]: \mathbb{R} \rightarrow \mathbb{Z}$ from the upper limit of the last summation denotes the ceiling function. $\square$

\section{MFT from Message Passing}
The iterator $\mathcal{G}_{\rm MF}: \Omega_0 \rightarrow \Omega_0$ only retrieves the $\hat{x}$-projected component $\underline{X}$ of mean spin polarization profile, whereas the $\underline{Y}$-component vanishes identically by the dimension reduction lemma. As a proof of principle to confirm validity of Lemma 1.1, we implement a standard message-passing algorithm~\cite{MessagePassing} from variational inference to alternatively recover the optimal MF probability measure without a priori deriving the form of the maximum entropy measure $\nu^{\star}$ in Eq.~\ref{eq:nu_star}. 

Let us revisit the dipolar system as a graphical model. The basic idea behind the MF message-passing algorithm is also to optimize the free energy functional $\mathcal{F}[\nu]$ over the space of measures $\nu = \prod_{i=1}^{n} \nu_i$ factorizable as product of the singleton functions $\nu_i(\vartheta)$. However, the optimization here is subject to the hard-coded constraints that $\nu_{i}$ on each node $i \in \mathcal{V}$ satisfies all the defining properties of a marginal probability measure, which leads to the update rule in the infinite-dimensional space $\mathcal{M}(\underline{\vartheta})$,
\begin{eqnarray}
    \begin{split}
        \hspace{-0.75 cm} {[\nu_{i}]}&_{\tau+1}(\vartheta)  \propto \\
        &\psi_{i}(\vartheta) 
        \prod_{j \in \mathcal{N}_i} \exp{\left[ - \beta \int d\vartheta' V_{ij}(\vartheta, \vartheta'){[\nu_{j}]}_{\tau}(\vartheta') \right]},
    \end{split}
\end{eqnarray}
where ${[\nu_i]}_{\tau}$ gives the marginal measure on vertex $i$ at the $\tau$th update, $\ln \psi_{i} = - \beta h_{i}$ captures the statistical weights of external fields, and $\mathcal{N}_i = \{ j \in \mathcal{V}: (i,j) \in \mathcal{E} \}$ denotes the neighborhood of vertices connected to the vertex $i \in \mathcal{V}$. The update can be viewed as a message-passing process on the factor graph $\Lambda$ with messages, $\Tilde{m}: \mathcal{E} \times [-\pi, \pi) \rightarrow \mathbb{R}^{+}$, passed back and forth along the edges,
\begin{eqnarray}
    &\hspace{-.5cm} \displaystyle \nu_{i}(\vartheta) \propto \psi_{i}(\vartheta) \prod_{j \in \mathcal{N}_{i}} \Tilde{m}_{ij}(\vartheta) & \mathcal{N}_{i} \mapsto i, \label{eq:MP_in}  \\
    &\hspace{-.5cm} \displaystyle \Tilde{m}_{ij}(\vartheta) \propto  \exp{\left[ -\beta \int d\vartheta' V_{ij}(\vartheta, \vartheta')\nu_{j}(\vartheta') \right]}  \hspace{0.25cm} & i \mapsto \mathcal{N}_{i}, \label{eq:MP_out}
\end{eqnarray}
where Eqs.~\ref{eq:MP_in} and \ref{eq:MP_out} indicate the flow of accessible information into and out of node $i$ respectively. In practice, we work with a dense but finite subset of the one-body phase space to approximate the marginal measures ${[\nu_{i}]}_{\tau}$ over the continuous variables $\vartheta_i$. Fig.~\ref{fig:BP_Iteration} below illustrates the performance of such a MF message-passing algorithm starting from a uniform prior ${[\nu_{i}]}_{0}(\vartheta) = 1/2\pi$. We adopt a discretized scheme with a finite sample of $10^2$ evenly spaced grid points on $[-\pi,\pi)$. A comparison between the MF profiles generated from the implicit message-passing ${[\nu_i]}_{\tau} \mapsto {[\nu_i]}_{\tau+1}$,
\smallskip
\begin{eqnarray}
    \underline{X}_{\tau}^{\nu} &=& \bigtimes_{i \in \mathcal{V}} \left\langle d_i\cos{\vartheta_i} \right\rangle_{ {[\nu_{i}]}_{\tau} } \mapsto \underline{X}_{\tau+1}^{\nu}, \\
    \underline{Y}_{\tau}^{\nu} &=& \bigtimes_{i \in \mathcal{V}} \left\langle d_i\sin{\vartheta_i} \right\rangle_{ {[\nu_{i}]}_{\tau} } \mapsto \underline{Y}_{\tau+1}^{\nu},
\end{eqnarray}
and the explicit recursion, $\underline{X}_{\tau} \mapsto \underline{X}_{\tau+1}$, reveals that $\mathcal{G}_{\rm MF}(\underline{X})$ indeed recovers the MF solution as claimed and meanwhile saves the computational time by orders of magnitude.
\bigskip

\begin{figure}[H]
\centering
\includegraphics[width=.4\textwidth]{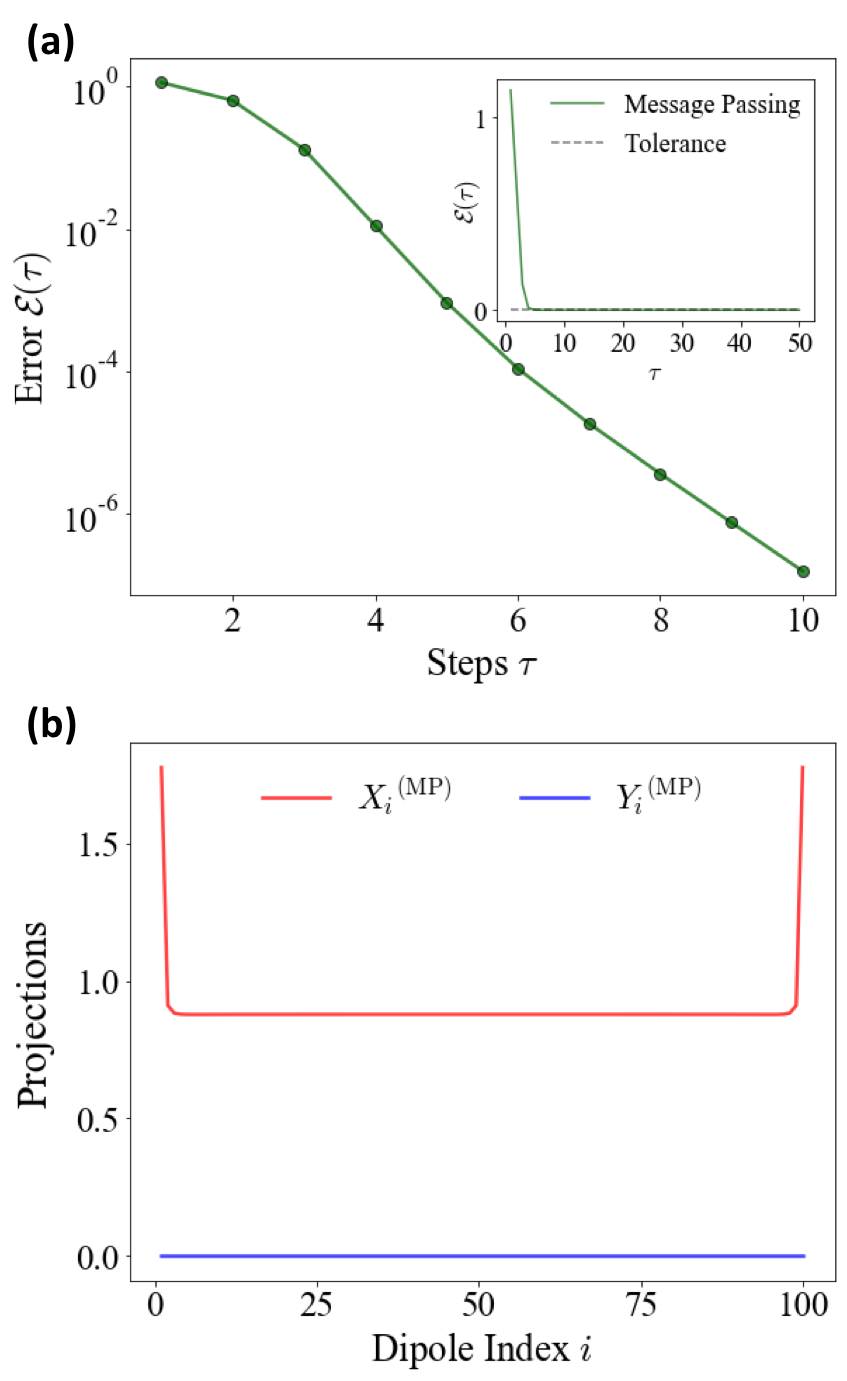}
\caption{Numerical closure of mean-field message passing. The algorithm is run for the model parameters $(n, \beta, E_{\rm ext}, d_{1}, d_{\rm bulk}, a) = (100, 1, 0.2, 2, 1, 1)$. $(a)$ Convergence error measured by successive mean discrepancy $(\underline{X}_{\tau+1}^{\nu}-\underline{X}_{\tau}^{\nu}, \underline{Y}_{\tau+1}^{\nu}-\underline{Y}^{\nu}_{\tau})$ is plotted over iterations, where the dashed line in black marks a reasonable tolerance.
$(b)$ Corresponding MF dipole configuration is plotted against the dipole positions on a 1D lattice.
}
\label{fig:BP_Iteration}
\end{figure}

\section{Higher Dimensional Spherical Spins}
The main convergence theorems, \textbf{Thm 2.1.} and \textbf{Thm 2.3.}, apply to spherical spin models where the orientational degrees of freedom $\underline{\omega}$ reside on a $(p-1)$-sphere $\mathbb{S}^{\hspace{0.03cm} p-1}$,
\begin{eqnarray}
    \hspace{-0.5cm} \vartheta_i \in [0,2\pi) \mapsto \omega_i \in
    \begin{cases}
    \mathbb{Z}_2 &~ p =1 \\
    [0,\pi]^{p-2} \times [0,2\pi) &~ p \geq 2
    \end{cases},
\end{eqnarray}
with system Hamiltonian, 
\begin{widetext}
\begin{eqnarray}
    \hspace{-0.75cm}  \mathcal{H}(\underline{\omega}|d_{\Lambda}) &=& \sum_{(i,j) \in \mathcal{E}} V_{ij}(\omega_i, \omega_j) + \sum_{i \in \mathcal{V}} h_i(\omega_i), \\
    &=& \frac{1}{2} \sum_{i \neq j=1}^{n} d_i d_j \hat{\mu}_i(\omega_i)^{\top}  \mathcal{T}_{ij} (r_{ij}) \hat{\mu}_j(\omega_j) - \sum_{i=1}^{n} d_i  \hat{\mu}_i(\omega_i)^{\top} \mathbf{E}^{\rm ext}_i,
\end{eqnarray}
\end{widetext}
on some fully connected graph $\Lambda = (\mathcal{V}, \mathcal{E})$, where $\mathcal{T}_{ij} \in \mathbb{R}^{p \times p}$ gives the two-body interaction disclosing a preference along, say, the primitive lattice vector $\hat{x} = (1,0,\cdots,0) \in \mathbb{R}^{p}$

Suppose $\mathcal{X}= V_{k}(\mathbb{R}^{p}) = \{ \mathcal{W} \in \mathbb{R}^{p \times k} : \mathcal{W}^{\top} \mathcal{W} = I_{k \times k} \}$, \textit{i.e.}, the set of orthonormal $k$-frames in $\mathbb{R}^{p}$. For example, consider
\begin{eqnarray}
    \hspace{-0.65cm} \mathcal{H}(\underline{\mathcal{W}}) = -\sum_{(i,j) \in \mathcal{E}} d_i d_j {\rm tr}\big( \mathcal{W}_i^{\top} \mathcal{W}_j \big) - \sum_{i \in \mathcal{V}} d_i {\rm tr}\big( E_i^{\top} \mathcal{W}_i \big),
    \label{eq:tetrahedral_SI}
\end{eqnarray}
where $\mathcal{W}_i \in V_{3}(\mathbb{R}^3) \equiv O(3)$ and $E_i \in \mathbb{R}^{3 \times 3}$. Under the MF assumption, the derived one-body measure with respect to the canonical Haar measure over $O(p)$, or more generally over $V_k(\mathbb{R}^p)$, takes the parametrized form,
\begin{eqnarray}
    \nu_i^{\star}(\mathcal{W}_i) = \frac{\exp\big[ \beta {\rm tr} \big( \Pi_i^{\top} \mathcal{W}_i \big) \big]}{_{0}F_1\big( p/2; \beta^2 \Pi_i^{\top} \Pi_i /4 \big)},
    \label{eq:Matrix_Fisher_SI}
\end{eqnarray}
where $_{0}F_{1}$ is the hypergeometric function
of matrix argument. The matrix parameter $\Pi_i = d_i U_i \gamma_i \in \mathbb{R}^{p \times k}$ under its polar decomposition of partial isometry $U_i$ and dilation $\gamma_i$ can be completely expressed in terms of the mean spin orientation $W_i = \mathbb{E}_{\nu_i^{\star}}[\mathcal{W}_i] \in \mathbb{B}_{p, k} = \{ \mathcal{W} \in \mathbb{R}^{p \times k} : {\rm tr} (\mathcal{W}^{\top} \mathcal{W} ) \leq k \}$. Assuming isotropy of the external field, $E_i \mapsto E_i I_{3 \times 3}$, the global maximizer $\underline{W}^{\ast} \in \mathbb{B}_{3,3}^n$ of the MF free energy functional must lie in the cone $PS_{+}^{3} = \{ \mathcal{W} = \mathcal{W}^{\top} \in \mathbb{R}^{3 \times 3} : x^{\top} \mathcal{W} x \geq 0, ~\forall x \in \mathbb{R}^3 \}$, based on the simple observations that ${\rm tr}(\gamma) \geq {\rm tr}(\Pi)$ for a polar decomposition $\Pi = U \gamma$, $\forall \Pi \in \mathbb{R}^{p \times p}$, and the measure $\nu_i^{\star}$ is invariant under the conjugacy $(\mathcal{W}_i, \Pi_i) \mapsto (g_1 \mathcal{W}_i g_2^{-1}, g_1 \Pi_i g_2^{-1})$, $\forall g_1 \in O(p)$ and $\forall g_2 \in O(k)$. In fact, a rearrangement argument~\cite{RearrangeIneq} shows that we should search for $\underline{W}^{\ast}$ within a subcone $D_{+}^{3}=\{ \mathcal{W} \in \mathbb{R}^{3 \times 3} : \mathcal{W}_{kl} = \kappa_l \delta_{kl}, \kappa_1 \geq \kappa_{2} \geq \kappa_{3} \geq 0 \}$ of diagonal matrices and introduce the projected MF coordinates,
\begin{eqnarray}
    \hspace{-0.5cm} \underline{X} = \begin{split}
        \big(d_1(W_1)_{11}, ~&d_1(W_1)_{22}, d_1(W_1)_{33}, \\
        &\cdots, \\
        d_n(W_n)_{11}, ~&d_n(W_n)_{22}, d_n(W_n)_{33} \big)
    \end{split} \in \bigtimes_{i \in \mathcal{V}} [0,d_i]^3,
\end{eqnarray}
where $(W_{i})_{kk} \leq 1$. Thus $ \mathcal{G}_{\rm MF} = \bigtimes_{i \in \mathcal{V}} d_i \mathcal{Q}(\beta d_i E_{{\rm MF},i})$ for which
\begin{eqnarray}
    E_{{\rm MF},i}(\underline{X};p) = E_i I_{p \times p} + \sum_{j \in \mathcal{V}}^{j \neq i} \sum_{I=1}^{p} (X_{j})_I e_I  e_I^{\top},
\end{eqnarray}
with rank-$1$ projectors $e_I e_I^{\top}$ defined by the standard basis $e_I$ of $\mathbb{R}^p$, and $\forall W \in D_{+}^{\hspace{0.025cm} p}$ we have
\begin{eqnarray}
    ~~\mathcal{Q}(W) = \left[- \nabla_{\underline{\kappa}} S_i \right]^{-1}(W_{11}, W_{22}, \cdots, W_{pp}),
\end{eqnarray}
where the one-body entropy $S_i$ of $\nu_i^{\star}$ from Eq.~\ref{eq:Matrix_Fisher_SI} depends on the $p$ singular values $\underline{\kappa} \in {\mathbb{R}^{+}}^{p}$ of the mean orientation due to the observed invariance under orthogonal conjugation (here $p = 3$). The iterator $\mathcal{G}_{\rm MF}$ picks up the previous form when we consider the simplified model with chirality frozen in Eq.~\ref{eq:tetrahedral_SI}, \textit{i.e.}, $\mathcal{W}_i \in V_2(\mathbb{R}^{3}) \cong SO(3)$. We view each $\mathcal{W} \in SO(3)$ as a rotation around some axis $\Vec{v} \in \mathbb{S}^{2}$ together with an angle $\vartheta \in [0,2\pi]$, and exploit the isomorphism $SO(3) \cong SU(2)/\mathbb{Z}_2$ using the adjoint representation,
\begin{eqnarray}
    \mathcal{W}(\vartheta, \Vec{v}) = \exp{\left[ \vartheta \Vec{v} \cdot \Vec{E} \right]}  \leftrightarrow \pm \exp{\left[\frac{i \vartheta \Vec{v} \cdot \Vec{\sigma}}{2} \right]},
    \label{eq:SO3_SU2}
\end{eqnarray}
where the rotation generators $\Vec{E}$ and Pauli matrices $\Vec{\sigma}$ form the standard basis of the Lie algebras $\mathfrak{so}(3)$ and $\mathfrak{su}(2)$ respectively. The above axis-angle information $(\vartheta, \Vec{v})$ is stored as a pair of antipodal vectors $\pm \hat{\mu} (\vartheta, \Vec{v}) \in \mathbb{S}^3 $ whose entries specify the RHS of Eq.~\ref{eq:SO3_SU2} via the identification,
\begin{eqnarray}
    \begin{split}
        \hspace{-0.75 cm} \pm \exp{\left[\frac{i \vartheta \Vec{v} \cdot \Vec{\sigma}}{2} \right]} & = \pm \begin{bmatrix}
        \phantom{-}b(\vartheta,\Vec{v})^{\phantom{*}} & c(\vartheta,\Vec{v})^{\phantom{*}} \\
        -c(\vartheta,\Vec{v})^{*} & b(\vartheta,\Vec{v})^{*}
        \end{bmatrix} \\
        &\hspace{-0.4 cm} \leftrightarrow \pm (\Re b, \Im b, \Re c, \Im c) =  \pm \hat{\mu}(\vartheta, \Vec{v}),
    \end{split}
\end{eqnarray}
where $\ast$ denotes the $\mathbb{C}$-conjugation. The tracial terms in the original Hamiltonian then appear as $\ell^2$-inner products, \textit{i.e.}, ${\rm tr}(\mathcal{W}_{i}^{\top} \mathcal{W}_{j}) = 4 ( \hat{\mu}_i^{\top} \hat{\mu}_j )^2-1$ and ${\rm tr}(E_{i}^{\top} \mathcal{W}_i) = E_i[ 4 (\hat{\mu}_i^{\top} \hat{x})^2-1 ]$, so we can write,
\begin{eqnarray}
    \begin{split}
        \hspace{-0.75cm} \mathcal{H}(\underline{\mathcal{W}}|d_{\Lambda}) \equiv \sum_{(i,j) \in \mathcal{E}} d_i d_j ( \hat{\mu}_i & \otimes \hat{\mu}_i )^{\top} \mathcal{T}^{ij} ( \hat{\mu}_j \otimes \hat{\mu}_j )  \\
        & -\sum_{i \in \mathcal{V}} d_i ( \hat{\mu}_i \otimes \hat{\mu}_i )^{\top} \mathbf{E}^{\rm ext}_i, 
    \end{split}
\end{eqnarray}
where $\mathcal{T}^{ij} = -2I_{4 \times 4} \otimes 2I_{4 \times 4}$ and $\mathbf{E}^{\rm ext}_i = E_i (2E_i^{1/2}\hat{x} \otimes 2E_{i}^{1/2}\hat{x})$. We have dropped a constant energy shift from the traces and restricted our attention to the diagonal subset of the product space $\mathbb{R}^4 \otimes \mathbb{R}^4$. Although the local effective field, $E_{\rm MF}$, gains a quadratic dependence on the MF coordinates, a conditioned version of Thm.2.3 applies under the high temperature or large external field limit upon accordant change of the convergence constants. On the other hand, both Thm.2.1 and Thm.2.3 hold if we replace the spin phase space $SO(3)$ with $SU(2)$ and matrix transpose with hermitian conjugate due to the diffeomorphism $SU(2) \cong \mathbb{S}^3$.

\section{Indirect Utility of MF Iterator}
We consider an indirect use of the iterator $\mathcal{G}_{\rm MF}$ to extract the MF solution when a particular model falls outside the model space region with direct iterator applicability. A hierarchy of iterator applicability is illustrated schematically in Fig.~\ref{fig:model_space} below. We recall that $\mathcal{G}_{\rm MF}$ establishes a deterministic walk in the space of configurational probabilities. For ferromagnetic spin models, this walk converges when the effective field $\mathbf{E}_{{\rm MF},i}(\underline{\theta})$ on each spin meets the positivity condition $\mathbf{E}_{{\rm MF},i}^{\top} \mathbf{E}_{{\rm MF},j} \geq 0$, $\forall (i,j) \in \mathcal{E}$ (irrespective of whether the converged spin polarity comes from optimal $\underline{\theta}$). Since we are interested in finding the global minimum of the MF landscape, now without precise resolution over the $\underline{\theta}$ coordinate, we employ a hybrid approach which relies on both deterministic and stochastic walk in search for optimal $\underline{\theta}$.

\begin{figure}[H]
\centering
\includegraphics[width=.5\textwidth]{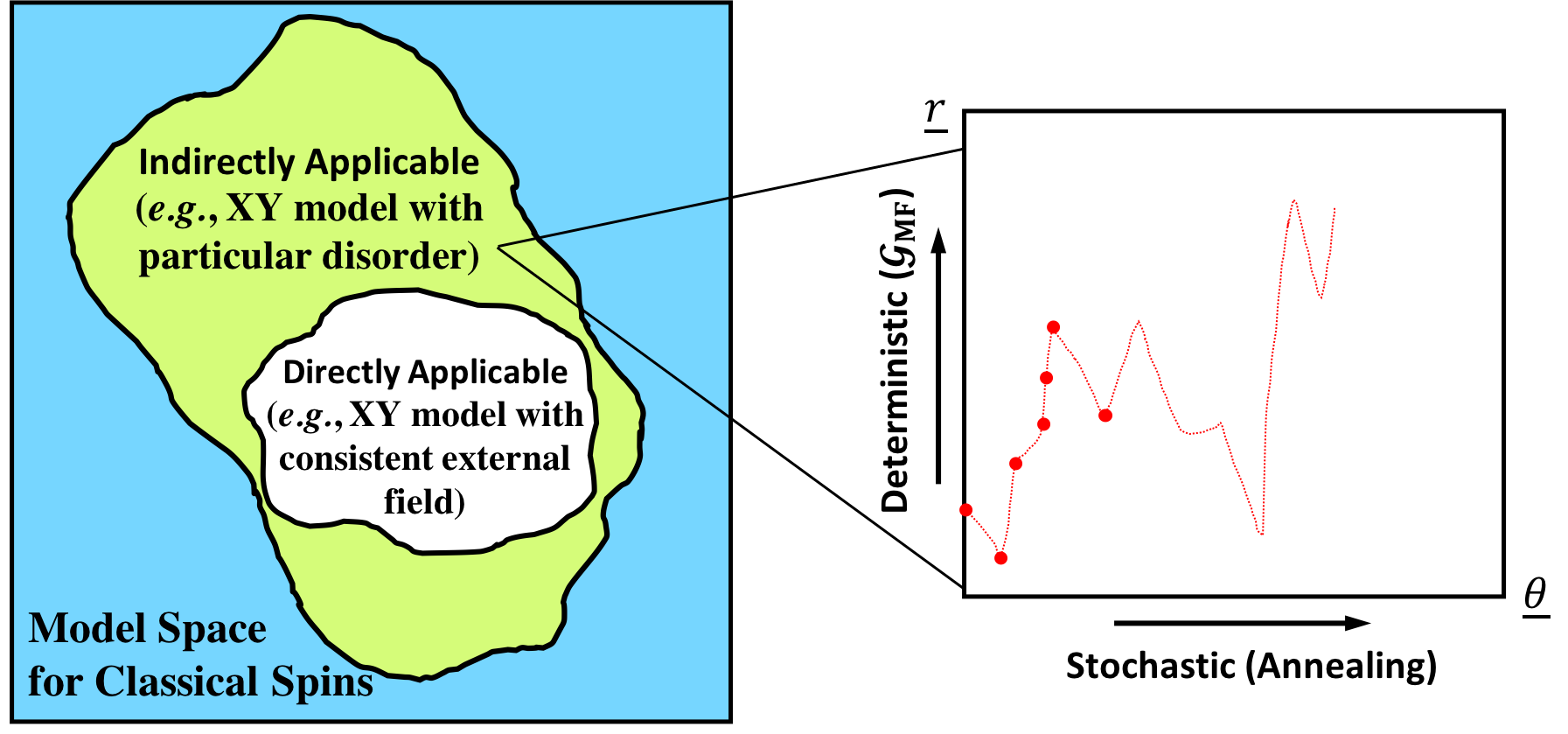}
\caption{Indirect use of MF iterator for solving spin models. The inset shows an example of hybrid approach that combines a deterministic walk from $\mathcal{G}_{\rm MF}$ at a given location on the MF landscape and a stochastic walk visiting different parts of the landscape. The red curve represents a generated stochastic trajectory whose evolution depends on evaluation of the MF free energy function $\mathcal{F}_{\rm MF}$ via $\mathcal{G}_{\rm MF}$, and the red dots correspond to states sampled along the trajectory.
}
\label{fig:model_space}
\end{figure}

In such scenarios, we can use $\mathcal{G}_{\rm MF}$ in junction with a global optimization routine to search across the MF landscape. Here, we choose the generalized stimulated annealing (GSA) schedule~\cite{GSA_1995, SA_2000} as our gradient-free optimization routine. Given some objective function $f: \Omega \subset \mathbb{R}^{n} \rightarrow \mathbb{R}$ over a search domain $\Omega$, GSA locates its global minimum by attempting stochastic moves in the search domain with a radial-symmetric visiting distribution,
\begin{eqnarray}
    p_{\rm trial}(\Delta \underline{x}, t) \propto \frac{T(t)^{-\frac{n}{3-q_{v}}}}{\left[ 1 + (q_{v}-1) \frac{\norm{\Delta \underline{x}}^2 }{T(t)^{ \frac{2}{3-q_v}} } \right]^{ \frac{1}{q_{v}-1} + \frac{n-1}{2} }} ,
\end{eqnarray}
where the visiting parameter $q_v \in (1,3]$ controls the shape of $p_{\rm trial}$ along a stochastic trajectory parametrized by the artificial time $t \in [1, \infty)$. Apart from setting the typical size of $\norm{\Delta \underline{x}}$, the artificial temperature, $T(t)$, also determines the likelihood of accepting the trial move $\underline{x} \mapsto \Delta \underline{x} + \underline{x}$ with a probability,
\begin{eqnarray}
    \hspace{-0.75 cm} p_{\rm accept}(\Delta\underline{x}, t;\underline{x}) = \min \left\{1, \left[1-\frac{\gamma(t) (1-q_a) \Delta f}{T(t)} \right]^{\frac{1}{1-q_a}} \right\},
\end{eqnarray}
where the accepting parameter $q_a$ controls the success of the trial move through the evaluated difference $\Delta f = f(\underline{x} + \Delta \underline{x}) - f(\underline{x})$, and the prefactor $\gamma(t)$ is commonly taken to be $\gamma = 1$ or $\gamma = t$ for reasonable convergence. To avoid the sign issue in the regime $q_{a} < 1$, we let $p_{\rm accept} = 0$ when $\gamma (1-q_a) \Delta f/ T < 1$. Over the course of optimum search, the annealing process occurs with continuously lowered temperature $T(t)$,
\begin{eqnarray}
    T(t) =  T(1) \frac{2^{q_{v}-1} - 1}{(1+t)^{ q_{v} - 1} -1 }.
\end{eqnarray}
Note that in the limit $(q_v, q_a) = (1,1)$, we essentially recover a Metropolis MCMC walk on the energy landscape $f(\underline{x})$, where a low energy state can be asymptotically reached.

To resolve the MF response in Fig.~\ref{fig:MCMC_XY}, we can identify our objective function, 
\begin{eqnarray}
    f(\underline{\theta}) = \mathcal{F}_{\rm MF} ( \underline{\theta}, \underline{r}^{c}(\underline{\theta}) ); ~~  \underline{r}^{c} =  \mathcal{G}_{\rm MF}(\underline{r}^{c}| \underline{\theta}),
\end{eqnarray}
on angular domain $\Omega = [-\pi/4, \pi/4]^{n}$, where $\underline{r}^{c}(\underline{\theta})$ represents the mean spin polarity satisfying the conditioned optimality. The rapid evaluation of the conditioned MF free energy $f(\underline{\theta})$ via a $\mathcal{G}_{\rm MF}$ iteration therefore provides the necessary ingredient for GSA calculation. For implementation of GSA, we use the optimize package available from SciPy.

\nocite{*}
\bibliography{main} 

\end{document}